# Visualizing the Hidden Architecture of Molecular Films with Phase-Resolved Rotational SFG Microscopy

Ben John, Tuhin Khan, Nasim Mirzajani, Martin Wolf, Alexander P. Fellows*, and Martin Thämer*

Fritz-Haber-Institut der Max-Planck-Gesellschaft, Faradayweg 4-6, 14195, Berlin, Germany

*Corresponding Authors

Email: fellows@fhi-berlin.mpg.de

Tel: +49 (0)30 8413 5140

Email: thaemer@fhi-berlin.mpg.de

Tel: +49 (0)30 8413 5145

## Abstract

The highly variable physico-chemical properties of thin molecular films play an essential role in numerous research fields ranging from biophysics to the fabrication of functional devices such as molecular sensors. The properties of molecular films are largely governed by their three-dimensional molecular structure which often exhibits important spatial heterogeneity, either naturally, or introduced deliberately. In order to understand and control these properties microscopic insight into structural parameters such as composition, molecular orientation and conformation, as well as molecular order is required, which, so far, represents a mostly unachieved experimental target. In this contribution we present a powerful experimental approach that can overcome this limitation. Using phase-resolved rotational sum-frequency generation (SFG) microscopy all of these structural parameters can be obtained with sub-monolayer sensitivity and at sub-micron resolution. In measurements of monolayer assemblies of mixed phospholipids, we uncover the molecular packing structure in previously-unattained detail and demonstrate the large potential of the technique for the elucidation of the complex architecture inside molecular films. The structural insight provided by this nonlinear microscopy approach spans all the way from the molecular to the macroscopic scale opening the door to a completely new type of interfacial studies.

# Introduction

Molecular assemblies at interfaces play a key role in a wide range of both natural and industrial processes. These include self-assembled systems such as biological membranes(*1*, *2*) and surfactant films(*3–6*), as well as other molecular boundaries, for example those at electrochemical interfaces(*7*, *8*) or even in tailor-made functionalized devices(*9*, *10*) (e.g., through molecular printing). Many of these systems are naturally or intentionally heterogeneous, with important spatial variations in their compositions, packing arrangements, and molecular conformations(*11*). These inhomogeneities typically lead to the formation of distinct regions with specialized physico-chemical properties that take on specific functions and are thus critical in determining the macroscopic behavior in biological and chemical processes. Examples of such specialized regions include the condensed domains in lipid membranes (lipid rafts)(*12*), electrochemically or catalytically active surface sites(*13*), as well as designed constructs with specific molecular sensors.(*14*, *15*) Gaining experimental access to the spatial distribution of the molecular structure and composition in such systems is therefore of great scientific and technological interest, but has nevertheless proven to be a difficult task. Major challenges range from very fundamental difficulties e.g. obtaining microscopy images of molecular monolayers, to more sophisticated endeavors such as identifying specific molecular species and characterizing the structural details of the molecular packing.

Over the past decades, several experimental techniques have been developed that can gain important insight into various aspects of the microscopic structure within molecular films. Among them are local probe techniques such as atomic force microscopy (AFM) (*16–19*) and scanning-tunneling microscopy(*20*, *21*) (STM), which can gain insight into topographical, electronic, and mechanical properties with nanoscale spatial resolution, as well as their recent advancements that also incorporate optical probes such as in AFM-IR(*22–29*) and scanning near-field optical microscopy(*30*, *31*) (SNOM) to provide chemical sensitivity. Beyond these, purely optical approaches like Brewster-Angle(*32–34*), ellipsometric(*35*, *36*), and phase-contrast microscopy(*37*, *38*) can provide morphological information, while Raman(*39*, *40*) and fluorescence microscopies(*41–44*) can add the molecular recognition, albeit usually requiring the introduction of specific tags. Finally, imaging methods based on higher energy light sources, such as photoemission electron microscopy (PEEM)(*45*, *46*), micro-grazing incidence x-ray diffraction (μ-GIXD)(*47*), and near-edge x-ray absorption fine structure (NEXAFS) microscopy(*48*) allow for elemental analysis along with some symmetry and structural information on the nano-to-microscale. Despite their combined benefits, however, gaining sensitivity to the intricacies of the molecular packing with these techniques is intrinsically restricted. Specifically, the structure in molecular films is characterized by how the molecules are individually oriented in space (i.e. direction and conformation), how they are packed relative to each other (i.e. their orientational distribution), along with any heterogeneity in composition and density. Although some of this information can be retrieved using the above techniques, some properties remain completely inaccessible (e.g. absolute directions and conformations) and the reconstruction of the desired structural picture down to the molecular level is not feasible. Without such a complete picture, the role that molecular structure plays in many macroscopic processes can only be partly addressed.

One technique that does have intrinsic sensitivity to absolute molecular orientations and their distribution is phase-resolved vibrational sum-frequency-generation (SFG) spectroscopy, which has become a popular and well-established characterization technique for studying interfaces and thin films.(*49–91*) One of the main strengths associated with this second-order vibrational probe is that it can correlate molecular recognition in a label free manner (via characteristic vibrational spectra) with conformational and orientational information, all with sub-monolayer sensitivity. This particularity of the technique is based on the unique connection between the sign of the nonlinear response and the absolute molecular orientation, forming peaks or dips in the resulting spectra.(*92–95*) An important consequence of this relation is that signals from isotropic media vanish (via cancellation), rendering SFG exclusively sensitive to the anisotropic structure. As most bulk media exhibit structural isotropy, SFG is hence often labelled an 'interface-selective' technique. While this interface selectivity is essentially irrelevant for most self-assembled or structured molecular films, the exclusivity to anisotropic structure nevertheless makes SFG a compelling method for their characterization. In particular, it means SFG can directly assess the degree of orientational ordering within the molecular film and, assuming the phase of the generated signal is determined (via heterodyne detection), also ascertain the absolute molecular orientations.(*96*)

More specifically, the generated signals in SFG experiments are governed by the second-order susceptibility $\chi^{(2)}$ – a rank-3 tensor that connects the polarization directions of all three interacting electromagnetic fields in all possible combinations. The orientational distribution of molecules within the measured sample spot is encoded in each of these tensor elements, but importantly with different functional forms. The combination of selected tensor components can therefore reveal the exact orientational characteristics of the molecules in space, described by parameters such as average tilt, azimuth and twist angles,(*97*) as well as width and shape of the orientational distribution. Importantly, this orientational information is obtained for the functional group that contains the specific vibrational mode being probed. Consequently, by combining such information from different bonds and functional groups within a molecule, not only the orientation of the whole molecule but also its exact conformation can in principle be obtained. A fundamental requirement of such a detailed structural analysis is, however, that individual tensor components (as many as possible) are isolated in a quantitative manner, which is usually done by performing multiple SFG spectroscopy measurements with varying beam polarizations combined with proper referencing. Based on these unique capabilities of the SFG technique in terms of structural characterization it seems very appealing to combine SFG spectroscopy with microscopic imaging. Such nonlinear microscopy approach would enable the structural elucidation to be transferred down to the microscopic scale, allowing the heterogeneity in these molecular films to be directly imaged.

Despite the enormous potential of SFG microscopy, it is nevertheless significantly less advanced than its spatially averaged spectroscopy counterpart, and far from an established technique.(*98–106*) In its mostly adopted spectroscopic implementation, the macroscopic averaging misses out on a great deal of the information in heterogeneous systems. Particularly, any structural distinctions between specialized surface regions become indistinguishable. Equally, at such a macroscopic scale, the majority of molecular films present effective in-plane isotropy, where any local in-plane directionality is cancelled in the average SFG response. While this 'ascent in symmetry' from the local to the spatially averaged structure still enables parameters such as the tilt angle to be calculated, disregarding the heterogeneity not only ignores a crucial aspect of the structure, but can even lead to misinformed conclusions.

The reason why the potential of SFG microscopy has so far not really been exploited is mainly related to technical difficulties. One of the most central obstacles is the fact that the SFG responses from molecular films are minute. While this restriction can be overcome in SFG spectroscopy due to the macroscopic integration of the responses, isolating the signal from a small surface region around 4-6 orders of magnitude smaller presents another level of difficulty. Therefore, while SFG microscopy has been successfully demonstrated for molecular samples,(*98*, *104*, *106–109*) this sensitivity challenge has so far placed significant restrictions on the types of samples that can be investigated along with the level of structural insight possible from them. For example, experiments have been performed on substantially thicker samples that also exhibit inherent symmetry breaking,(*102*, *110*) thus yielding much larger responses from a given surface region. This, however, effectively replaces the in-plane integration that is inherent to SFG spectroscopy with out-of-plane integration over the entire film thickness.

SFG microscopy has also been applied to much thinner films even down to molecular monolayers, but these studies have primarily been on metallic substrates, exploiting the significant field enhancement normal to the interface that results from surface plasmon resonances and image dipoles to overcome the sensitivity limitation.(*104*, *108*, *111*, *112*) These studies, however, yield relatively little insight into the molecular structure as only a single element of the second-order susceptibility, $\chi^{(2)}_{ZZZ}$ (normal to the surface), gives an appreciable signal due to in-plane field cancellation by the metal. While this still allows the heterogeneity across the surface to be imaged, such a limitation hinders the structural analysis that requires the combination of multiple susceptibility elements (as mentioned above). Ultimately, for many heterogeneous samples that also display substantial structural anisotropy, the restriction to just a single tensor element means a great deal of the rich structural information contained within $\chi^{(2)}$ is inaccessible. Until recently,(*113–118*) hyperspectral SFG imaging with sub-monolayer sensitivity on a dielectric substrate had not been demonstrated, let alone with phase resolution to ascertain absolute orientations or polarization dependent studies. Even more importantly, the quantitative isolation of multiple tensor components from SFG microscopy measurements of a sample has so far never been achieved — preventing the desired comprehensive structural analysis described above.

Here, we present our recently developed phase-resolved SFG microscope that overcomes the limitations. In rotational microscopy measurements on mixed phospholipid monolayers, we demonstrate the wide-ranging capabilities of this method and gain unprecedented insight into the highly heterogeneous and, so far, little known molecular structure that is formed in these model-membranes. This is achieved through several technical and methodological advances, that combine i) a wide-field imaging approach with interferometric (heterodyned) time-domain scanning with ii) a revised imaging geometry and iii) a newly developed *paired-pixel balanced imaging* system that enables trivial alignment, free choice of polarizations, and substantially improved signal-to-noise ratios,(*119*) as well as iv) employing an azimuthal scanning scheme at different polarization combinations, that allows for fully deconvoluting the SFG responses into their different tensorial contributions based on a Fourier decomposition. Using this approach, we demonstrate for the first time quantitative imaging of multiple isolated $\chi^{(2)}$ tensor elements from a molecular monolayer. Comparisons of these individual components in their absolute units then allow us to determine structural parameters such as the tilt angle, degree of in-plane packing order, and molecular densities / composition, all completely quantitatively and with full spatial resolution.

This work provides a comprehensive description of this vibrational microscopy approach, giving full technical and experimental details, describing the complete theoretical framework behind the quantitative isolation of different $\chi^{(2)}$ tensor elements from rotational SFG microscopy measurements, and showing how detailed molecular structure information can be determined based on these experiments. Overall, this demonstration of the capabilities of rotational SFG microscopy highlights its potential to become an established and powerful characterization technique in the molecular sciences by providing previously inaccessible structural insight. Importantly, the technique does not require specialized sample conditions – it can be performed under ambient conditions and at buried interfaces and doesn't require any labels or local probes. Furthermore, as SFG is widely applicable to any systems displaying a lack of centrosymmetry, many of the presented concepts can be extended to other sample systems such as van der Waals heterostructures,(*116*) ordered antiferromagnetic materials,(*117*) or even for histological imaging of whole cells or biological tissues.(*120–123*)

## SFG Microscope and Balanced Imaging

### *Spectral Acquisition*

Vibrational sum-frequency signals are generated by nonlinear frequency mixing between two laser pulses, an infrared beam that probes vibrational resonances and a visible upconversion beam that is typically tuned to be far off any sample resonance. The resulting nonlinear signal contains the desired vibrational information that can be represented in the form of a second-order vibrational spectrum. In SFG microscopy, the desired data set is thus three dimensional: two dimensions representing spatial axes (the image), and the third being the vibrational frequency axis. Several approaches exist to acquire such data, employing different measurement concepts. Such methods include widefield narrowband techniques, where spatial SFG images are captured on a CCD camera while stepping through vibrational frequencies using a narrowband IR laser.(*124*, *125*) Alternatively, some SFG microscopes use broadband infrared pulses in combination with narrowband upconversion pulses.(*104*, *126*) This technique allows for simultaneous detection of SFG signals from all vibrational frequencies within the IR bandwidth. However, to obtain the desired frequency resolution, the use of a polychromator is required, which is incompatible with widefield imaging (as one dimension of the 2D detector is used to map the different frequencies). In consequence, this approach requires lateral rastering of the sample using a confocal microscope.

The microscope presented here utilizes an alternative strategy: widefield imaging with broadband infrared pulses in a time-domain scanning scheme. Spectral acquisition occurs interferometrically by adding an SFG reference pulse (local oscillator, LO) and modulating the time delays between the different input laser pulses using a home-built nonlinear interferometer. A clear benefit of such a time-domain approach is that it directly yields both spectral phase and amplitude of the nonlinear responses and that it can easily be combined with widefield imaging. Furthermore, this approach allows for the straightforward acquisition of linear optical microscopy images by illuminating the sample only with the LO, as well as phase-resolved nonlinear (SFG) images using all input pulses. This feature bears important practical advantages, as discussed later. The employed measurement concept is adapted from our time domain SFG spectrometer that is described in detail elsewhere.(*127*) Here we only briefly discuss its basic principle.

Tunable mid-infrared and visible upconversion pulses independently enter the interferometer, with the IR being split into two portions. The first, smaller portion (~10%) is collinearly coupled with the visible beam and focused onto a nonlinear crystal (z-cut quartz) to generate the LO. The resulting beam (containing visible and LO) is then passed through a delay stage and collinearly coupled with the second, larger portion (~90%) of the IR to generate a single collinear output containing the three broadband pulses. A schematic of the interferometer is shown in the top-left of Figure 1.

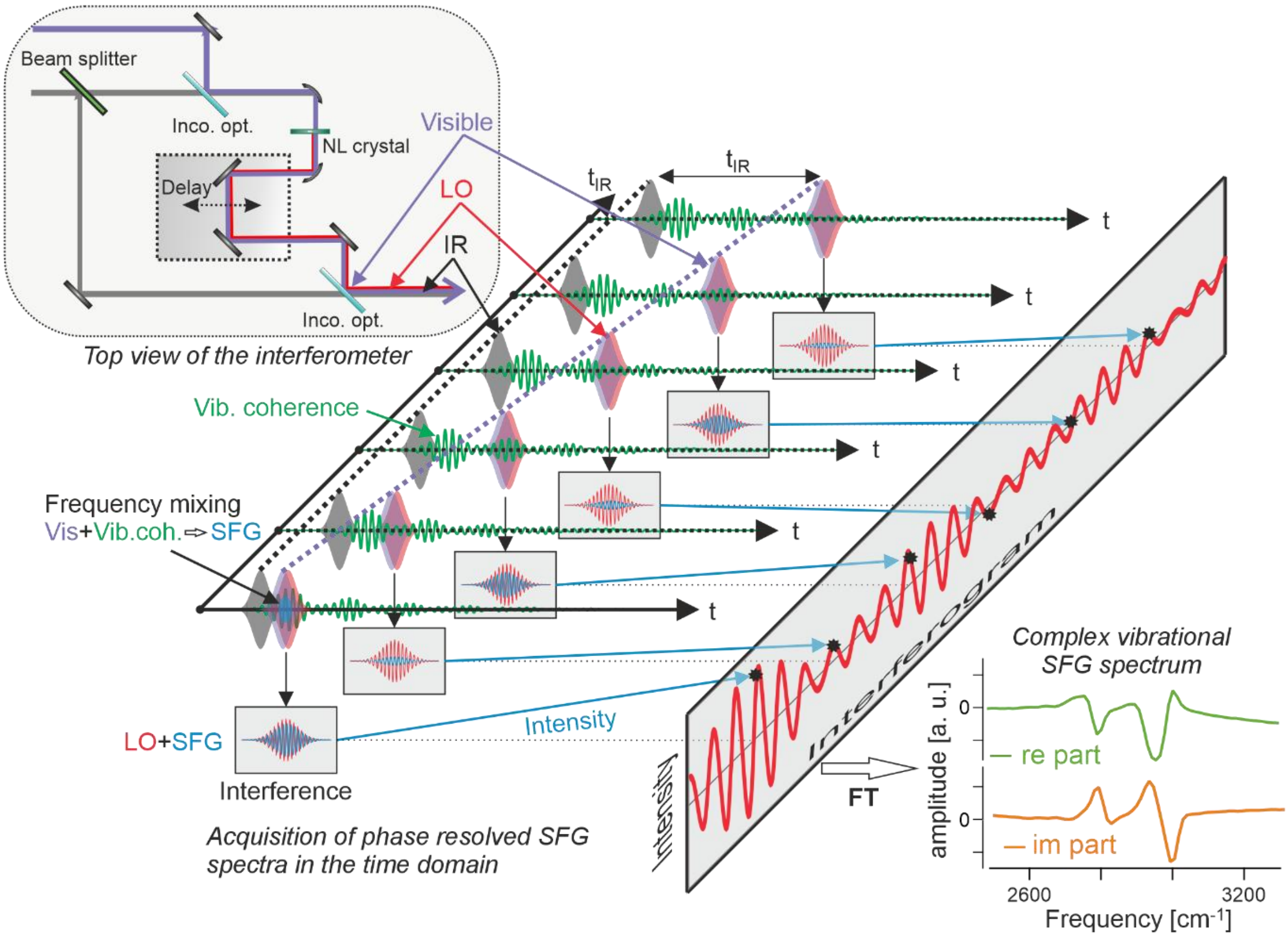

*Figure 1:* ***Time domain interferometric approach to SFG Microscopy****. A top-down schematic of the interferometer is shown in the top-left, accompanied by a flow diagram illustrating the generation of an interferogram. This occurs through the heterodyning interference between the SFG signal (blue) and the 'reference' local oscillator (LO, red). As the time-difference ($t_{IR}$) between the IR (gray) and the combination of visible upconversion (purple) and LO is increased, the phase and amplitude of the SFG is modulated, reflecting the vibrational coherence. This modulation is thus imprinted onto the heterodyned SFG signal. Fourier transforming (FT) the resulting interferogram results in a complex SFG spectrum, as depicted on the right side.*

The collinear output of the interferometer is then focused onto a sample where the infrared and the visible upconversion pulses generate a sum-frequency signal that interferes with the LO. The resulting reflected heterodyned response is collected by an objective and imaged onto a camera, which allows the individual microscopic SFG responses within the irradiated area to be spatially resolved. By scanning the time delay between the IR and the other two pulses using the interferometer stage, the phase and amplitude of the SFG is modulated with respect to the LO such that their interference generates an interferogram, as depicted in Figure 1. By imaging this interference as a function of the delay, isolated interferograms are thus generated from each detected pixel. A Fourier transform (FT) of the delay axis then converts this dataset into complex-valued 3D hyperspectral images, which are normalized in phase and amplitude using the data from a separate SFG microscopy measurement of a reference sample (z-cut quartz).

***Imaging Geometry and Balanced Detection***

To accomplish the desired widefield imaging, a reflective Schwarzschild objective is used, where the sample irradiation occurs through a custom hole drilled through the objective (as illustrated in Figure 2). This configuration allows for oblique sample irradiation while maintaining distortion-free widefield imaging.(*119*) The oblique incidence beam geometry facilitates access to all polarization combinations — a crucial requirement for fully characterizing the orientations of molecular species, as mentioned above. To generate a broad illumination spot, a relatively soft focus is chosen using a long focal length off-axis parabolic mirror. The resulting SFG signal from the sample is collected by the objective, along with the reflected LO, and, after removal of the fundamental pump frequencies using a set of spectral filters, the entire beam (containing the LO and the SFG signal) is imaged onto a camera using a tube lens. To achieve low-noise data acquisition, our recently developed *paired-pixel balanced imaging* technique is employed.(*119*) As per typical balanced detection, the polarization of the LO is chosen to be perpendicular to the desired polarization direction of the generated SFG response. Subsequently, both the LO and SFG polarizations are rotated by 45° using an achromatic waveplate (WP) and the entire beam is then transmitted through a polarizing beam splitter (PBS), which divides it into two segments with horizontal and vertical polarizations. This process of polarization mixing brings the SFG signal and the LO into interference in both segments. As this is done in a widefield imaging approach, it generates two full images on the camera (A and B in Figure 2). If the sample is irradiated only by the LO (linear optical microscopy mode) the two images are identical replicas (i.e. $A = B$). When interfering with the SFG signal, however, the interference terms have opposite sign, which makes the two images deviate ($A \neq B$). The desired image containing only the interference term can then be isolated by taking the difference between the two images. In this balanced output, the noise originating from laser intensity and pointing fluctuations is highly suppressed, leading to typical enhancements in signal-to-noise up to a factor of 10. Furthermore, by isolating the interference term, only the signal portions that are coherent with the LO contribute to the resulting hyperspectral images, i.e. only the desired sample SFG signal is obtained. Any parasitic stray light or unwanted higher order nonlinear signal contributions are suppressed by this coherence filter.

Such a balanced detection scheme in combination with widefield imaging, however, requires a precise match of corresponding pixels between the two recorded images so they can be correctly overlapped. This is important to avoid degradation of the spatial resolution when forming the difference image, as well as to ensure efficient noise cancellation. As shown in our previous work, only corresponding pixel pairs present equal intensity noise traces, and consequently only these pixel pairs are well balanced.(*119*) The required map of corresponding pixel pairs can be determined using the microscope in its linear optical mode (LO only) in combination with a resolution test target (e.g. USAF-1951) as shown in Figure 2. Since both recorded images must be exactly equal, the task is simply a precise matching of the observed fine structures of the test target. This is done based on a 2D convolution between the two images. Importantly, the determined map of correlated pixels only depends on the alignment of the optical components after the sample (i.e. between the sample and the camera) and is therefore independent of sample alignment and even fairly insensitive to changes in pump beam pointing. In consequence, this map of correlated pixels can be generated once after the initial alignment of the microscope and used for all subsequent sample measurements. The same measurement is also used to correctly determine the focus, which can easily be found based on the linear optical image of the test target. Because of the oblique sample irradiation, the position of the illuminated sample region and thus the center of the illuminated area on the camera sensitively depends on the exact distance between the sample and the objective. Any sample can therefore be precisely placed into the focal plane by tuning the sample-objective distance until the center of the illuminated area appears in the correct position on the camera screen. This always ensures a precise sample alignment independent of the specific sample properties.

***Background Suppression***

One important challenge in nonlinear experiments when employing a collinear beam geometry is the suppression of parasitic SFG signals. Since all beams (including infrared and upconversion) overlap spatially on the optics throughout the setup, SFG signals can, in principle, be generated at their surfaces if they also overlap temporally. These parasitic contributions will then appear as a non-zero background in the final spectra. To achieve high precision and reliability in the obtained results they should therefore be fully suppressed in the measurement. Such suppression is achieved in the presented microscope by manipulating the relative timings of the different laser pulses so that they only overlap spatially, but not temporally on the optics before the sample.(*128*, *129*) This is feasible here as the pulses are all femtosecond (broadband) pulses, so only moderate delays are required, which can be attained by introducing a delay optic just before the sample, namely a LiF window (DO, Figure 2) at normal incidence into the pump beams. The dispersion in LiF retards the infrared pulse with respect to the visible upconversion pulse ($\Delta t_A$, Figure 2), with the result that the delay stage position that corresponds to the temporal overlap at the sample is sufficiently far from temporal overlap for the other optics in the setup (and thus they do not generate any parasitic signals). In the context of the time domain scanning approach, this means that the interferogram containing the parasitic signal contribution is displaced to negative time delays with respect to the sample interferogram and is thus efficiently suppressed in the measured data.(*128*, *129*)

The downside of introducing the LiF window in the beam is that its dispersion also acts on the LO that now lags behind the upconversion pulse, and thus also lagging behind the SFG signal generated from the sample. This timing mismatch ($\Delta t_B$, Figure 2) would thus highly reduce the amplitude of the sample interferogram if not accounted for. To compensate for such a delay, we exploit the orthogonal polarizations of LO and SFG signal, which allows their relative delay to be tuned using a birefringent crystal mounted in the correct orientation behind the objective (delay compensation optics, DCO).(*128*, *129*) A schematic representation of this temporal delay procedure is shown in Figure 2.

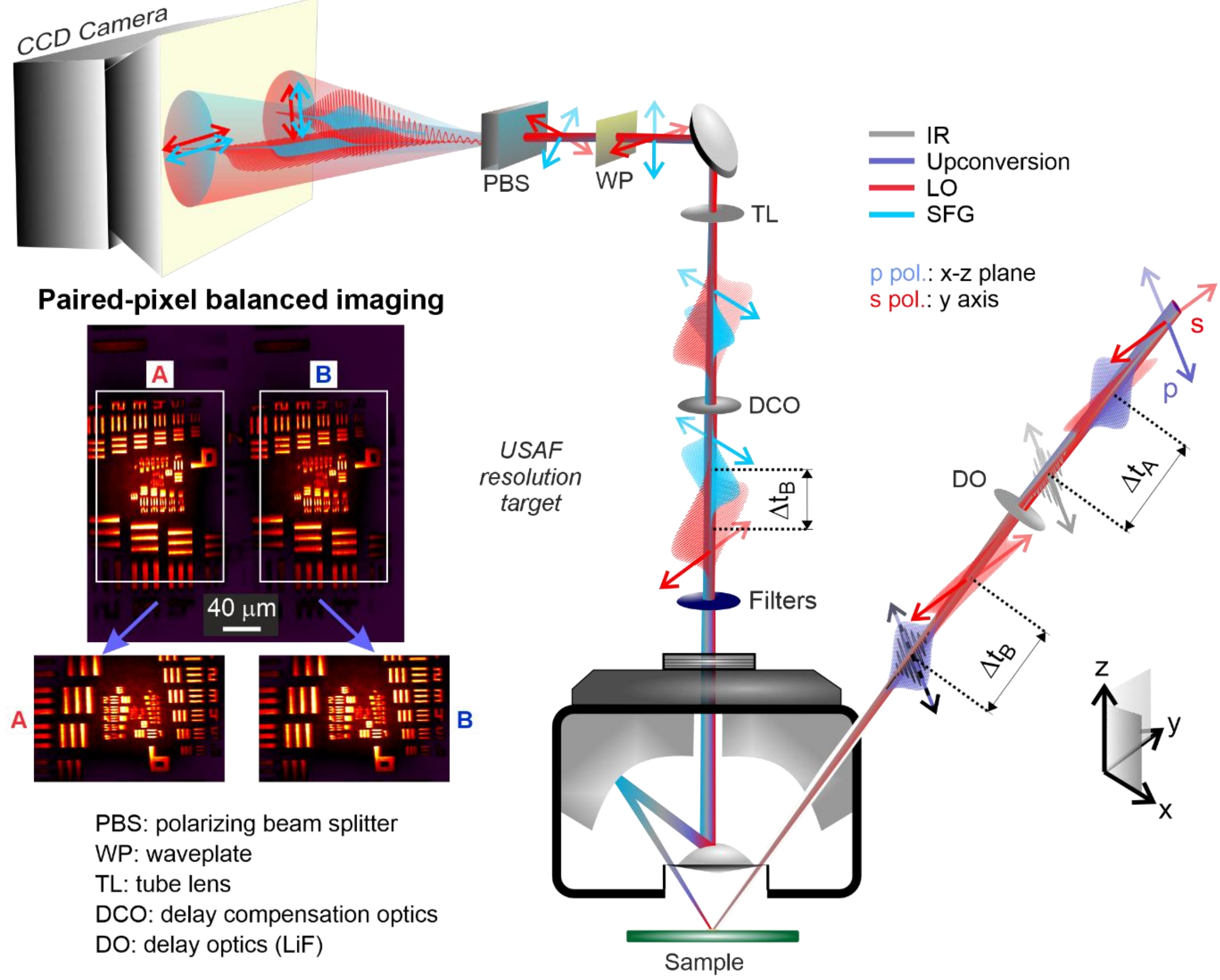

*Figure 2:* ***Schematic of the SFG microscope and paired-pixel balanced imaging.*** *The IR (gray), visible upconversion (purple), and LO (red) beams are collinearly directed through a custom-made hole in a reflective objective. The resulting SFG signal (blue) and reflected incident beams are collected in a wide-field manner and subsequently pass through frequency-selective filters to isolate the SFG and LO. Delay compensation optics (DCO) are then used to temporally overlap the two pulses, followed by them being focused using a tube lens. Before arriving at the camera, a series of polarization optics are used to perform paired-pixel balanced detection, whereby two images (A and B) are created on the same camera with opposite sign interference.*

## Microscopy Studies of Phospholipid Assemblies

Having established the experimental concepts behind the phase-resolved SFG microscope, including its balanced imaging procedure, time domain scanning approach, and parasitic signal suppression, we proceed to demonstrate its performance and capabilities. While structural heterogeneity is generally rife in molecular films and often plays the critical role in defining the macroscopic functionality of the system, this study focusses on one specific class, namely phospholipid membranes. Here, we investigate monolayers containing a mixture of saturated (DPPC) and unsaturated (POPC) lipids (Figure 3a). Upon modifying the surface density, these lipids form highly heterogeneous molecular assemblies, characterized by liquid condensed (LC) domains surrounded by a liquid expanded (LE) phase (see Figure 3a)(*130*). These phospholipid films are crucial in biology as the fundamental building blocks of cell membranes, where they assemble into bilayers with two oppositely oriented monolayer leaflets. Despite their significance, the molecular structures within these inhomogeneous layers are not well understood, even for the comparatively simplistic case of a two-component monolayer, modelling a single leaflet of the membrane. Important questions regarding the heterogeneous molecular structure in such systems remain unanswered, such as the precise compositions of the LC and LE phases, as well as the distribution of molecular orientations within their specific packing arrangements. While it is evident that all phospholipid molecules align mostly parallel to each other with their hydrocarbon tails pointing generally upwards, the conformation and specific tilt angles of the tail-groups within each phase, as well as the extent of any in-plane packing order, remains essentially unknown.

In this study, a mixture of DPPC and POPC in a 4:1 ratio is compressed in a Langmuir trough and immobilized by casting onto a fused silica substrate using Langmuir-Blodgett deposition. To enhance the contrast and enable an unambiguous identification of the spectral contributions from the two phospholipids, POPC is fully deuterated (dPOPC). That way the DPPC signals become the sole contributors to the C-H stretching region (2800 - 3000 $cm^{-1}$), with the dPOPC resonances being shifted to the C-D stretching region (2050 - 2250 $cm^{-1}$).

***Single Pixel Sub-Monolayer Sensitivity***

In a first experiment, full hyperspectral SFG microscopy images of the sample in the C-H stretching range are measured in the PPP polarization scheme (both pump beams and the detected SFG signals are P-polarized, LO is S-polarized, following the requirements for balanced detection where the LO is orthogonal to the SFG). Figure 3b shows a subset of the measured raw data, namely the unbalanced 'A' and 'B' images at different time delays ($t_{IR}$), as well as the corresponding balanced interferometric response (A-B). Figure 3c then shows the corresponding Fourier-transformed and referenced response in the form of the spectral imaginary parts at three selected frequencies. SFG spectra of such phospholipid assemblies are typically dominated by the vibrational response of the terminal $CH_3$ groups of the aliphatic chains (symmetric stretch (SS) at 2880 $cm^{-1}$, Fermi resonance (FR) at 2940 $cm^{-1}$, and antisymmetric stretch (AS) at 2960 $cm^{-1}$), while the corresponding $CH_2$ signals are largely suppressed due to cancellation. The selected three frequencies correspond to the $CH_3$ SS and AS responses ($\omega_2$ and $\omega_3$, respectively), along with a fully non-resonant image at 2820 $cm^{-1}$ ($\omega_1$), which is included for comparison. The two resonant images clearly show the presence of roughly circular regions of ~10 µm diameter that have increased signal amplitudes, corresponding to the LC domains that are formed in the phospholipid monolayer. In contrast, the nonresonant image does not show any spatial contrast and essentially represents the noise floor.

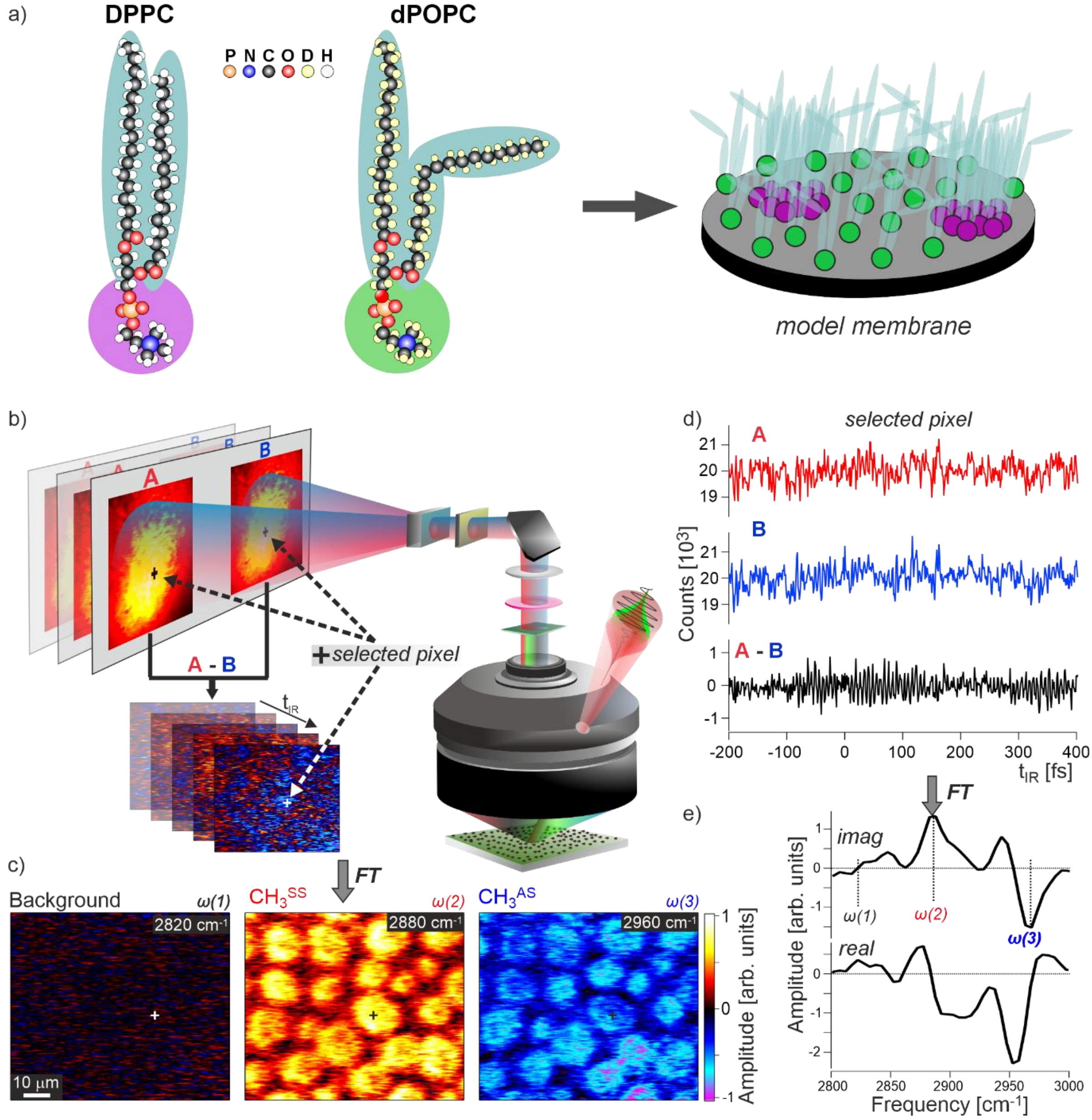

*Figure 3:* ***Paired-pixel balanced imaging of a molecular monolayer.*** *(a) Schematic of the structure of the protonated DPPC and deuterated dPOPC lipids. These lipid molecules are compressed using a Langmuir trough to form a simplistic model membrane that exhibits phase-separated liquid condensed (LC) domains surrounded by a liquid expanded (LE) phase. (b) Schematic of the microscope, illustrating its geometry and the balanced imaging setup. The balancing of the two images generates a time trace image. (c) Three different frequency planes of the Fourier-transformed response, showing the imaginary amplitudes at 2820, 2880, and 2960 cm*$^{-1}$*, corresponding to an off-resonant background, the resonance of the* $CH_3$ *symmetric stretch, and the resonance of the* $CH_3$ *antisymmetric stretch, respectively. (d) Time traces of a selected single pixel from each image (A and B), along with its corresponding balanced subtraction (A-B). Note the difference is presented on a different y-axis scale. (e) Fourier transform of the balanced time trace from (d), showing the resulting complex SFG spectrum from a single pixel.*

This result demonstrates that such condensed domains in a molecular monolayer can indeed be visualized in excellent quality by the SFG microscope. To further evaluate its performance, we assess the signal-to-noise ratio apparent in a single pixel. Figure 3d shows the raw interferometric time traces of the corresponding pixel in each unbalanced image (A and B) along with the balanced result (A-B). On comparing the time traces, it becomes evident that the unbalanced raw traces are dominated by noise, while an interferogram just becomes visible in the balanced trace. This enhancement in the signal-to-noise ratio by roughly a factor of 10 (*119*) is the key for obtaining the presented high-quality images. Figure 3e then presents the resulting Fourier-transformed spectrum of the DPPC phospholipids that are located in the selected sample pixel. As expected, the spectrum is dominated by the vibrational responses of the terminal $CH_3$ group (i.e. those at $\omega_2$ and $\omega_3$ as well as the FR at 2940 $cm^{-1}$). It does, however, also show small but non-negligible contributions from $CH_2$ stretches at ~2840 and 2900 $cm^{-1}$, indicating incomplete cancellation of these responses.

The caliber of the obtained spectrum is remarkable considering that it is obtained from a single pixel, which corresponds to just ~$2\cdot10^5$ DPPC molecules (see Supplementary Information). At first sight, this number might not seem particularly impressive when compared to other optical microscopy techniques such as Fluorescence or tip/surface-enhanced Raman (TERS/SERS) that can readily achieve sensitivity even down to the single molecule level. However, for a fair comparison, three things need to be taken into account: Firstly, we do not only *detect* the molecular species, we also obtain fully phase-resolved SFG spectra from these local molecular assemblies, which are label-free and uniquely contain the desired additional structural information such as molecular orientations and order. Secondly, the fact that C-H vibrations have much smaller hyperpolarizabilities than e.g. carbonyl vibrations clearly suggests that the achieved sensitivity of our microscope would also allow for detecting the signals from much smaller numbers of molecules if samples with stronger resonators were being probed, as is typically the case in fluorescence microscopy or Raman techniques. Finally, due to the presented widefield approach we obtain this information for more than 100,000 different sample locations simultaneously and without the need for any field enhancement effects.

***Azimuth- and Polarization-Dependent Imaging***

The presented measurement in Figure 3 serves as a proof of principle experiment demonstrating the achieved sensitivity of SFG microscopy. In the following, we extend these studies to analyze the molecular structures that are formed in these phospholipid monolayers. The focus therefore lies on the question to what extent the LC domains differ from the LE phase in the 3D orientation of their constituent molecules, their degree of in-plane ordering, as well as their overall density and chemical composition.

Access to these structural quantities can be obtained from the second-order susceptibility, $\chi^{(2)}$, which is the material-specific constant connecting the amplitudes of the input fields ($E_{VIS}$ and $E_{IR}$) with the measured output SFG field ($E_{SFG}$), as in Eq. 1.

$$E_{SFG} = \chi^{(2)} E_{VIS} E_{IR} \tag{1}$$

This complex-valued quantity is a rank-3 tensor containing 27 potentially unique non-zero elements and is typically specified in the laboratory frame coordinate system, $\{x, y, z\}$, which is defined such that the incident laser beams are contained within the $xz$-plane and the $z$-axis is normal to the sample surface. Importantly, as this sample property describes the macroscopic second-order response of the system, it doesn't represent the molecular response, which is instead determined by the hyperpolarizability tensor of the individual molecule, $\beta$. For obvious reasons, $\beta$ is customarily placed in a different coordinate system, namely that corresponding to the molecular frame, $\{a, b, c\}$. In order to connect these two quantities, two things need to happen: the molecular responses need to be transformed into the laboratory frame, and they need to be spatially averaged (i.e. over each individual pixel in microscopy). This process is described in Eq. 2,

$$\chi_{ijk}^{(2)} = \rho \sum_{p,q,r} \langle R_{ip} \cdot R_{jq} \cdot R_{kr} \rangle \beta_{pqr} \qquad (2)$$

where $\langle \cdot \rangle$ represents macroscopic averaging, $\rho$ is the molecular surface density, and $R$ represents a coordinate transformation matrix, typically given in terms of three Euler angles: $\{\theta, \phi, \psi\}$, with $\theta$ being the tilt angle from the surface normal, $\phi$ the in-plane azimuth, and $\psi$ the twist angle about the main molecular axis (defined as the $c$-axis). Eq. 2 thus shows how the aforementioned structural information is contained within the SFG response—its dependence on $\rho$ can be used to extract the density and composition, while the three Euler angles describe the 3D molecular conformation, and any reduction in signal from the macroscopic averaging speaks to the orientational order.

While $\chi^{(2)}$ contains all of this rich structural information, the relationship between the SFG responses and the molecular structure is not quite so straightforward. There are four main reasons for this:

1. The triple product of the Euler transformations results in the different orientational parameters ($\theta$, $\phi$, and $\psi$) becoming significantly convoluted and not generally separable.
2. The macroscopic averaging or symmetry of different functional groups can render the response independent on certain orientational parameters, such that they cannot be isolated. A clear example of this is the in-plane azimuth direction being isotropically distributed if averaged over a large enough sample area.
3. The SFG response is measured in a specific polarization combination, which generally combines multiple tensor elements. For example, as P-polarized light is defined with its electric field in the $xz$-plane, the PPP polarization combines $\chi^{(2)}$ elements with any combination of $x$ or $z$ indices—meaning a total of 8 contributions.
4. The measured response is influenced by experimental parameters such as the incidence angle(s), which determine the amount of $x$ and $z$ contributions probed by P-polarized light, and Fresnel factors, which relate the incident and measured fields to those 'felt' locally by the molecules.

Overall, this means the measured SFG signals are given by a so-called effective susceptibility, $\chi_{eff}^{(2)}$, that is defined for a particular polarization combination and contains all of the experimental factors. An example of this is shown in Eq. 3 for the PPP polarization combination, which has even been already simplified due to the collinear beam geometry we adopt (all beams have the same incidence angle to the sample, $\theta'$) resulting in the contributions from four of the eight tensor components cancelling with each other.(*113*)

$$\begin{aligned}\chi_{eff,PPP}^{(2)} = &- \cos^3\theta' \, L_x^3 L_x^2 L_x^1 \chi_{xxx}^{(2)} - \sin\theta' \cos^2\theta' \, L_x^3 L_x^2 L_z^1 \chi_{xxz}^{(2)} \\ &+ \sin^2\theta' \cos\theta' \, L_z^3 L_z^2 L_x^1 \chi_{zzx}^{(2)} + \sin^3\theta' \, L_z^3 L_z^2 L_z^1 \chi_{zzz}^{(2)}\end{aligned} \qquad (3)$$

Here $L_p^n$ is the nonlinear Fresnel factor along the $p$-axis for the frequency $\omega_n$. Eq. 3 shows that the effective response involves the sum of multiple individual tensor elements, which are differently dependent on the molecular orientational parameters, macroscopically averaged, and scaled by unique experimental factors. This emphasizes that the general relationship between the measured SFG responses and the desired structural parameters is extraordinarily complicated, and extracting them is no mean feat.

While the expressions for the effective response, such as that in Eq. 3, represent the generalized expression, they can often be simplified based on structural assumptions, such as those given above in point 2. If the macroscopic averaging leads to effective in-plane isotropy, this unfortunately means no information on the in-plane orientation can be obtained, but it also removes any dependency on this parameter and results in the cancellation of certain tensor elements e.g. the $\chi_{xxx}^{(2)}$ and $\chi_{zzx}^{(2)}$ components in Eq. 3. By further isolating specific resonances that are independent on the twist angle (e.g. the methyl symmetric stretch), the entire expression can be substantially simplified to only contain one or perhaps two tensor elements, and only be a function of the molecular tilt angle. This basis is often the foundation of many SFG spectroscopy studies, where the signals are spatially integrated over a large sample area, allowing for the determination of the average tilt angle by comparing different polarization combinations. By reducing the spatial integration in microscopic investigations, however, the assumption of in-plane isotropy is often invalid. In fact, for the mixed phospholipid systems investigated here, we have categorically demonstrated this in previous work simply by probing in the SSS polarization combination, which only probes the $\chi_{yyy}^{(2)}$ tensor element and should therefore vanish under the conditions of in-plane isotropy.(*92*, *95*, *114*) The SSS measurements revealed significant contributions from the LC phase domains, highlighting the substantial in-plane anisotropy at the single pixel level. The presence of this significant in-plane anisotropy therefore means such simplifications cannot be made, complicating the relation between the SFG signals and the molecular structure, but importantly also paves the way to far greater structural insight, for example by characterizing the orientational distribution within the local packing structure.

A powerful method to disentangle the structural parameters from the measured SFG signals is to combine the SFG hyperspectral imaging with azimuthal scanning, i.e. recording the SFG signals as a function of the sample rotation. This is achieved simply by mounting the sample within a motorized rotation stage with its rotation axis aligned to the principal axis of the objective. By rotating the sample, we modulate the azimuthal Euler angle while leaving the other structural parameters unchanged. Taking a Fourier transform of the rotational dependence—converting to rotational (azimuthal) frequencies—means the different tensor elements are separated based on their dependence on $\phi$. Before the set of hyperspectral images for the different rotations can be Fourier-transformed, however, the corresponding pixel for each rotation must obviously be precisely overlapped which is done by back-rotation of the different images and subsequent lateral matching.

Considering the effective PPP response, as shown in Eq. 3, each component takes the form of a triple product of the spatial '$x$' and '$z$' coordinates. By rotating the sample by $\varphi$, $z$ will be left unchanged, whereas $x$ will follow $\cos\varphi$. Therefore, each component will follow a different function of $\cos\varphi$ up to cubic dependence. Specifically, $\chi_{zzz}^{(2)}$ will transform as 1, $\chi_{zzx}^{(2)}$ as $\cos\varphi$, $\chi_{xxz}^{(2)}$ as $\cos^2\varphi$, and $\chi_{xxx}^{(2)}$ as $\cos^3\varphi$. Under the trigonometric relations, the $\chi_{zzz}^{(2)}$ component will only appear in the 0-fold rotational frequency, the $\chi^{(2)}_{zzx}$ will only contribute to the 1-fold response, $\chi^{(2)}_{xxz}$ to the 0-fold and 2-fold, and $\chi^{(2)}_{xxx}$ to both the 1-fold and 3-fold rotational frequencies. Measuring SFG images under azimuthal scanning thus offers a clear route to deconvoluting the different tensor elements, and thus simplifying the analysis for extracting structural parameters. The result of such a measurement is a 4-dimensional matrix where 2 dimensions (x, y) represent the spatial coordinates in the image while the 3rd and 4th dimension correspond to vibrational- and rotational frequency (ω, $f_n$), respectively.

At this point, it is important to briefly discuss the nature of the susceptibility contributions that are obtained from the rotational Fourier transformation, which have the general form $\chi^{(2)}_{ijk}(f_n)$. In contrast to the "regular" tensor components of the form $\chi^{(2)}_{ijk}$ that describe contributions to the nonlinear response for one specific sample rotation, $\chi^{(2)}_{ijk}(f_n)$ contains information on all sample rotations within the x-y plane. It still preserves vibrational frequency dependence of the response in form of a complex spectrum but additionally includes a rotational phase for $n > 0$. As a result, $\chi^{(2)}_{ijk}(f_n)$ is doubly complex – containing both a spectral and rotational phase. In simple terms that means that one can in principle extract for each mode in the spectrum its amplitude and its direction in the x-y plane. Mathematically this can be described as in Eq. 4.

$$\chi^{(2)}_{ijk}(\omega, f_n) = \sum_m \langle \frac{A^m_{ijk}}{\omega - \omega_m + i\Gamma_m} \cdot g^n_{ijk}(\theta_m, \psi_m) \cdot e^{in\phi_m} \rangle \quad (4)$$

Here, the first term in the summation product describes the complex Lorentzian line-shape of a specific vibrational mode, *m*, the second term, $g^n_{ijk}(\theta_m, \psi_m)$, is a function of the orientation of the functional group in $\theta_m$ and $\psi_m$ (and therefore mode-dependent), and the final term, $e^{in\phi_m}$, represents the azimuth-dependence as a complex rotational phase, with the ⟨ ⟩ denoting the macroscopic averaging over the probed area (as in Eq. 2).

The rotational Fourier transform thus separates-out the azimuthal dependency, resulting in a complex vibrational spectrum (product of the first two terms), which is independent on the azimuthal rotation, multiplied by a complex rotational phase factor that shows the orientation of each mode $m$ (e.g. the different functional groups) in the x-y plane. In consequence, the rotational Fourier components of certain tensor elements become closely related e.g. $\chi^{(2)}_{yyz}(f_0) = \chi^{(2)}_{xxz}(f_0)$ and $\chi^{(2)}_{yyy}(f_1) = \chi^{(2)}_{xxx}(f_1) \cdot e^{i\frac{\pi}{2}}$.

Based on this procedure, different tensor components can be isolated and analyzed. For PPP, however, the four contributions are still intermixed in the different rotational frequencies (see Table 1). While it may seem that it is possible to disentangle them using all four of the rotational harmonics (0- to 3-fold), this requires prior knowledge of the in-plane distribution as the higher harmonics are increasingly sensitive to the degree of in-plane order.(*118*) Nevertheless, given that the $\chi^{(2)}_{xxz}(f_0)$ and $\chi^{(2)}_{xxx}(f_1)$ components are directly analogous to $\chi^{(2)}_{yyz}(f_0)$ and $\chi^{(2)}_{yyy}(f_1)$, respectively, combining the PPP measurements with those in the SSP and SSS polarization combinations allows for a full deconvolution. Table 1 shows a summary of the contributions to the 0- and 1-fold harmonics in each of the PPP, SSP, and SSS polarization combinations.

|  | PPP | SSP | SSS |
|---|---|---|---|
| 0-Fold | $-\sin\theta'\cos^2\theta'\, L_x^3 L_x^2 L_z^1 \chi_{xxz}^{(2)}(f_0)$ $+\sin^3\theta'\, L_z^3 L_z^2 L_z^1 \chi_{zzz}^{(2)}(f_0)$ | $\cos\theta'\, L_y^3 L_y^2 L_z^1 \chi_{yyz}^{(2)}(f_0)$ | - |
| 1-Fold | $-\cos^3\theta'\, L_x^3 L_x^2 L_x^1 \chi_{xxx}^{(2)}(f_1)$ $+\sin^2\theta'\cos\theta'\, L_z^3 L_z^2 L_x^1 \chi_{zzx}^{(2)}(f_1)$ | $\sin\theta'\, L_y^3 L_y^2 L_x^1 \chi_{yyx}^{(2)}(f_1)$ | $L_y^3 L_y^2 L_y^1 \chi_{yyy}^{(2)}(f_1)$ |

*Table 1: Summary of the contributions from different tensor elements to each of the PPP, SSP, and SSS polarization combinations in the 0-fold and 1-fold rotational frequency harmonics.*

As an aside, for the case of in-plane isotropic systems only the 0-fold responses contribute to the SFG signal while all other Fourier components vanish. This property together with the resulting equality of the $\chi_{yyz}^{(2)}$ and $\chi_{xxz}^{(2)}$ tensor components has widely been exploited in previous SFG spectroscopy investigations that compare PPP and SSP spectra to calculate the average tilt angle. Along these lines it is important to note that the $\chi_{yyz}^{(2)}(f_0)$ and $\chi_{xxz}^{(2)}(f_0)$ responses for the general symmetry case are precisely those that would be measured for the case of in-plane isotropic systems. Therefore, by performing the rotational-dependent measurements, and subsequent isolation of the isotropic ($f_0$) components, the same comparisons can be expanded to systems not possessing in-plane isotropy.

The application of this approach to the phospholipid monolayers is presented in Figure 4, which shows the azimuthal-dependent hyperspectral data in each polarization combination, where the recorded SFG responses have been referenced by z-cut quartz and converted into absolute units for comparability. As discussed above, the data sets behind each polarization combination and rotational frequency are complex valued three-dimensional (x, y, ω) matrices, making it difficult to represent this data graphically in a compact form. In this particular study, however, a convenient reduction in data dimensionality is possible by performing a singular value decomposition (SVD) of all spectra within each individual hyperspectral image. As shown in the Supplementary Information, the spectral data is highly dominated by the 1st SVD component, which, to a good approximation, allows the data to be represented by i) a normalized complex spectrum (the respective 1st SVD component), ii) a magnitude image (showing the contribution of the spectrum to each individual pixel (x, y) in absolute units), and iii) a phase image that shows for $f_n > 0$ the x ,y direction in which the spectrum reaches its maximum. Furthermore, for better clarity, only the imaginary parts of the spectra (absorptive line shapes) are plotted and for better comparability all amplitude images are represented on the same scalebar. The respective phase images for the 1-fold responses are shown in the Supplementary Information.

The top-half of Figure 4 shows the 0-fold rotational frequency components, where the heterogeneous phase structure is clearly present in the PPP and SSP images, while the SSS image only shows noise. This is expected as the SSS polarization should only yield in-plane contributions that are modulated by the rotation, thus yielding no 0-fold response. As such, the normalized spectrum for SSS, despite indicating it might contain resonant information, is meaningless since it effectively has no intensity in the image. The bottom-half of Figure 4 then shows the 1-fold rotational harmonic data. Here, the rotational magnitude images (which are now independent of the specific Euler angle, $\phi$) also show significant signals, although are generally weaker than the 0-fold responses and now only show the LC domains, with the response in the LE phase being at the level of the noise floor. Unlike in the 0-fold, the SSS does show significant intensity in the 1-fold contribution, as expected. While these data already reveal numerous qualitative information on the molecular structure of the investigated sample a precise analysis at this point is hampered by the different experimental factors (Fresnel and beam geometry, see Supplementary Information) that are still contained within the effective susceptibilities.

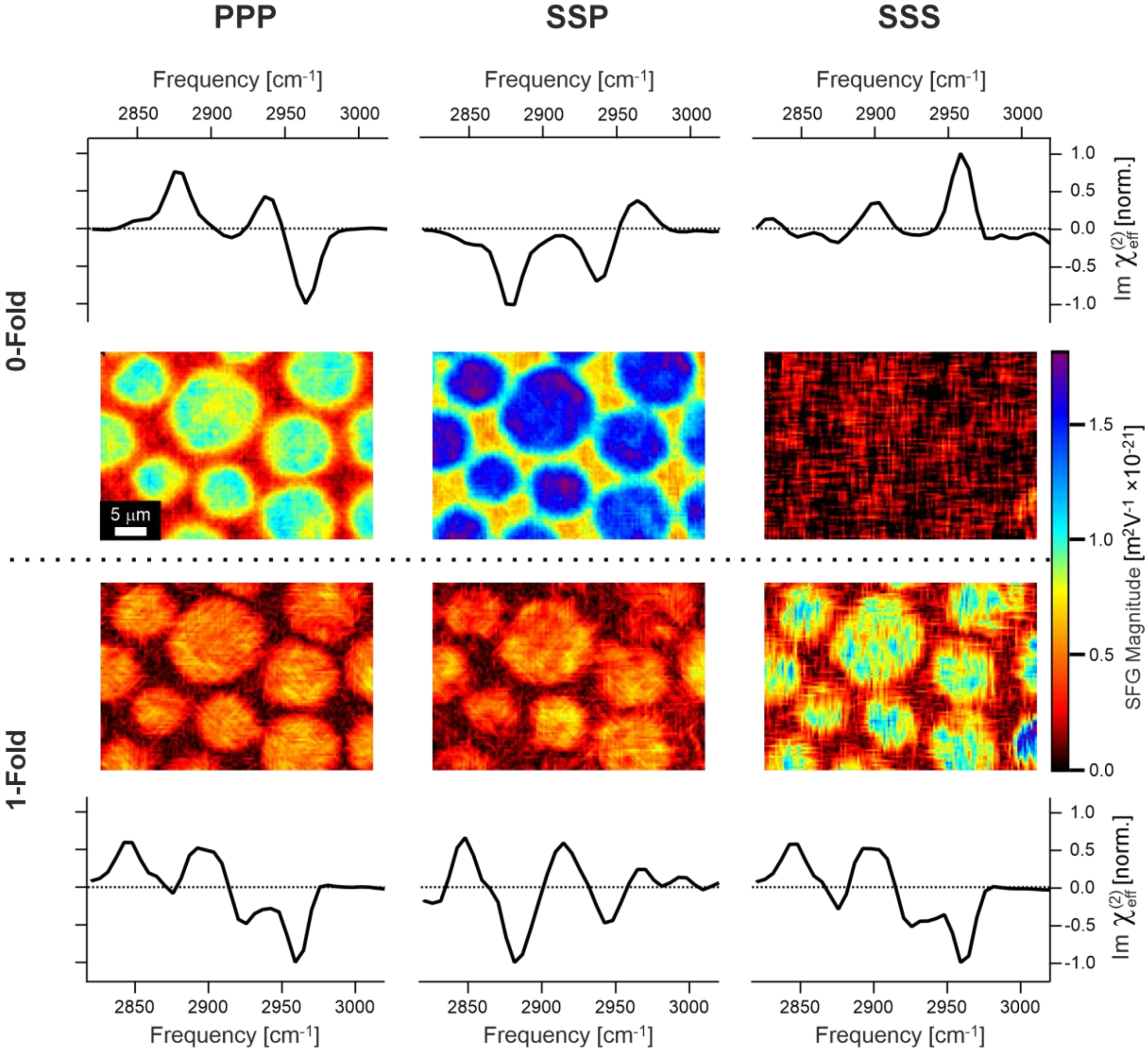

*Figure 4:* ***Polarization- and azimuth-dependent SFG hyperspectral imaging.*** *Spectra and images in the 0-fold and 1-fold rotational frequency components of three distinct polarization combinations: PPP, SSP, and SSS, in the C-H stretching region. The spectra are shown only through their imaginary parts and have been normalized to their strongest peak (to either 1 or -1), with the corresponding absolute amplitudes of the effective second-order susceptibilities,* $\chi_{eff}^{(2)}$*, being reflected in the rotational magnitude images (all plotted on the same scale for clarity). Note: the Fresnel factors and beam geometry factors are still included in these magnitudes.*

***Isolation of Tensor Elements***

With the different rotational frequency harmonics of the three polarization combinations in hand, we can now turn to the isolation of the individual tensor elements. Isolating the $\chi^{(2)}_{xxx}(f_1)\left(=\chi^{(2)}_{yyy}(f_1)e^{i\frac{\pi}{2}}\right)$, $\chi^{(2)}_{xxz}(f_0)\left(=\chi^{(2)}_{yyz}(f_0)\right)$, and $\chi^{(2)}_{yyx}(f_1)$ elements is rather trivial since they are the sole contributions to the SSS 1-fold, SSP 0-fold, and SSP 1-fold harmonics, respectively. They simply need to be corrected for their associated experimental factors (see Supplementary Information). Once these components have been isolated, extracting the $\chi^{(2)}_{zzx}(f_1)$ and $\chi^{(2)}_{zzz}(f_0)$ contributions from PPP also becomes relatively straightforward. The procedure for each component is outlined in Eqs. 5-9 and schematically shown in Figure 5a.

$$\chi^{(2)}_{xxx}(f_1)=\chi^{(2)}_{yyy}(f_1)e^{i\frac{\pi}{2}}=\frac{\chi^{(2)}_{SSS,eff}(f_1)}{L^3_yL^2_yL^1_y}e^{i\frac{\pi}{2}} \tag{5}$$

$$\chi^{(2)}_{yyx}(f_1)=\frac{\chi^{(2)}_{SSP,eff}(f_1)}{L^3_yL^2_yL^1_x} \tag{6}$$

$$\chi^{(2)}_{xxz}(f_0)=\chi^{(2)}_{yyz}(f_0)=\frac{\chi^{(2)}_{SSP,eff}(f_0)}{\cos\theta'\,L^3_yL^2_yL^1_z} \tag{7}$$

$$\chi^{(2)}_{zzx}(f_1)=\frac{\chi^{(2)}_{PPP,eff}(f_1)+\cos^3\theta'\,L^3_xL^2_xL^1_x\chi^{(2)}_{xxx}(f_1)}{\sin^2\theta'\cos\theta'\,L^3_zL^2_zL^1_x} \tag{8}$$

$$\chi^{(2)}_{zzz}(f_0)=\frac{\chi^{(2)}_{PPP,eff}(f_0)+\sin\theta'\cos^2\theta'\,L^3_xL^2_xL^1_z\chi^{(2)}_{xxz}(f_0)}{\sin^3\theta'\,L^3_zL^2_zL^1_z} \tag{9}$$

The resulting isolated individual tensor elements are then shown in Figure 5b, again represented with normalized spectra and the 2D images containing the absolute magnitudes (once again, calculated using SVD). Furthermore, the contributions from different vibrational modes to the spectra are determined by a multi-Lorentzian fit based on a global fitting procedure of the entire set of complex spectra (see Supplementary Information for details). The raw data are shown as points and the fitted spectra as solid lines, with their individual contributions shown with dashed lines and shading—where $CH_3$ resonances are highlighted in blue and $CH_2$ resonances in green. Overall, this global fitting yields exceptional overlap with the raw data and gives resonance peaks and dips at the expected known frequencies of aliphatic C-H resonances.

Comparing the obtained magnitude images it first becomes evident that the $\chi^{(2)}_{zzz}(f_0)$ and $\chi^{(2)}_{zzx}(f_1)$ contributions contain significantly more apparent noise than $\chi^{(2)}_{yyx}(f_1)$, $\chi^{(2)}_{xxz}(f_0)$ and $\chi^{(2)}_{xxx}(f_1)$. There are two reasons for this: Firstly, as shown in Eqs. 5-9 above, the $\chi^{(2)}_{yyx}(f_1)$, $\chi^{(2)}_{xxz}(f_0)$, and $\chi^{(2)}_{xxx}(f_1)$ elements are extracted from a single measurement, whereas the $\chi^{(2)}_{zzx}(f_1)$ and $\chi^{(2)}_{zzz}(f_0)$ are calculated from a combination of two, and thus the noise is compounded. Furthermore, and perhaps more importantly, the experimental scaling pre-factors for $\chi^{(2)}_{zzx}(f_1)$ and $\chi^{(2)}_{zzz}(f_0)$ are comparatively small (approximately 3 times smaller than the corresponding pre-factors for $\chi^{(2)}_{yyy}(f_1)$ and $\chi^{(2)}_{yyz}(f_0)$, see Supplementary Information). The division by these factors in Eqs. 5-9 hence results in the noise being significantly enhanced in $\chi^{(2)}_{zzx}(f_1)$ and $\chi^{(2)}_{zzz}(f_0)$ (cf. $\chi^{(2)}_{yyx}(f_1)$, $\chi^{(2)}_{xxz}(f_0)$, and $\chi^{(2)}_{xxx}(f_1)$). Nevertheless, the $\chi^{(2)}_{zzz}(f_0)$ contribution still shows clear contrast with strong signals in both phases, and, on close inspection, the $\chi^{(2)}_{zzx}(f_1)$ also shows slight contrast, with larger signals in the LC phase. Overall, the data quality is by far good enough for a detailed analysis and, more importantly, as these data are now independent on any experimental factors, they represent the pure material properties and the results can thus directly be interpreted in terms of molecular structure.

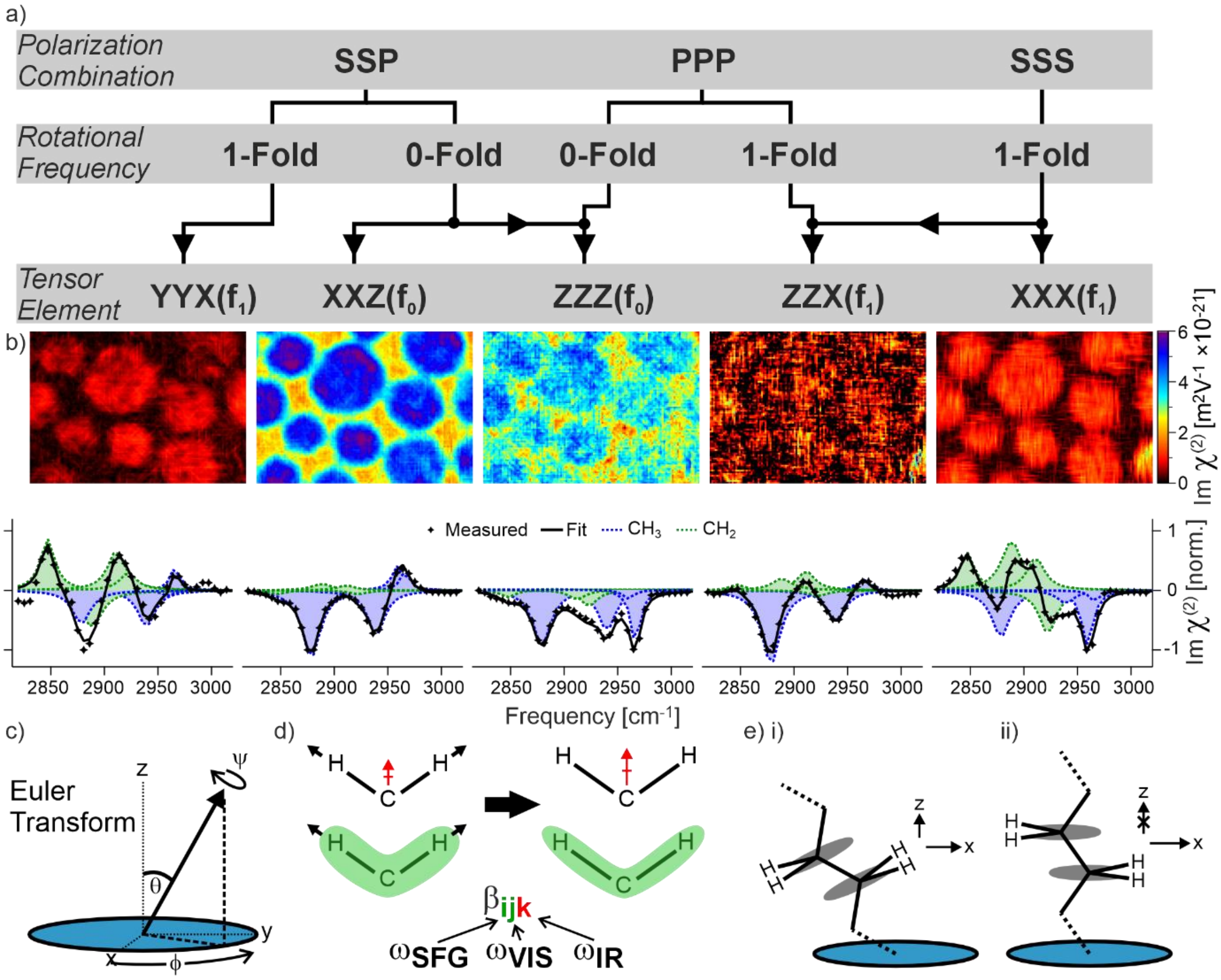

*Figure 5:* ***Extraction of individual tensor elements.*** *(a) Schematic workflow showing how the SFG signals in different polarization combinations and their resulting rotational frequency harmonics are combined to isolate the different tensor elements. (b) Images and spectra showing the hyperspectral magnitudes of the five different tensor elements in absolute units. The spectra are shown through their imaginary parts, having been normalized to their maximum peak (to 1 or -1), with the magnitudes of the susceptibility elements being reflected in the 2D images. All images are plotted on the same color scale for clarity. The spectra are also shown with their contributing resonances ($CH_3$ in blue and $CH_2$ in green), as determined from a global fitting procedure (see Experimental section for details). (c) A schematic showing the Euler transform between the molecular and lab frames. (d) The $CH_2$ symmetric stretch, showing the change in dipole and polarizability during the vibrational coordinate, and how these parameters are encoded in the hyperpolarizability, $\beta$, elements. (e) Diagrams depicting the orientation of $CH_2$ groups for different tilts of the lipid tails, showing (i) $CH_2$ groups with significant z-character for a tilted chain and (ii) $CH_2$ groups essentially in-plane for upright tails.*

The data show strong contributions from the $\chi^{(2)}_{xxz}(f_0)$ and $\chi^{(2)}_{zzz}(f_0)$ tensor elements, which are spectrally dominated by the resonances of the terminal $CH_3$ group of the two aliphatic chains of the DPPC molecule. A clear contrast between LC and the LE phases can be seen with generally larger signals from DPPC in the LC domains. However, as there is also substantial signal present in the LE phase it is clear that DPPC is present in both phases but likely in different densities. Compared to the $\chi^{(2)}_{xxz}(f_0)$ and $\chi^{(2)}_{zzz}(f_0)$ tensor elements, the signals in the $\chi^{(2)}_{yyx}(f_1)$ and $\chi^{(2)}_{xxx}(f_1)$ components are distincly smaller but still show substantial amplitude in the regions of the LC domains. The presence of these 1-fold signals unambiguously shows that, inside the LC domains, the DPPC molecules form anisotropic structures in the x-y plane (parallel to the interface), which means that they must possess preferential in-plane orientation within the sub-micron sample area of each pixel. In contrast, within the LE phase, the $\chi^{(2)}_{yyx}(f_1)$ and $\chi^{(2)}_{xxx}(f_1)$ signals are essentially zero, meaning the DPPC molecules in the LE phase must be isotropically distributed without any in-plane ordering.

The analysis of the corresponding spectra further reveals that, besides the $CH_3$ responses, the $\chi^{(2)}_{yyx}(f_1)$ and $\chi^{(2)}_{xxx}(f_1)$ components contain comparatively large contributions from $CH_2$ groups, here especially evidenced by the strong symmetric stretch vibration of aliphatic $CH_2$ at ~2840 $cm^{-1}$. Interestingly, this $CH_2$ signal does not show considerable contributions to the $\chi^{(2)}_{xxz}(f_0)$ and $\chi^{(2)}_{zzz}(f_0)$ components, showing that the tilt angle of the contributing $CH_2$ groups must be close to 90 degrees. Moreover, it is remarkable that the 2840 $cm^{-1}$ mode does not contribute to the $\chi^{(2)}_{zzx}(f_1)$ tensor element although this component includes a vibrational excitation along the x-axis. This observation can be rationalized by analyzing the symmetry of the $CH_2$ group. The $CH_2$ moiety can be treated as essentially flat, with both its symmetric and antisymmetric stretches being dominated by hyperpolarizability contributions in the plane of the atoms. This arises as the dipole is contained within this plane throughout the vibrations (i.e. is always zero across the bonds), and the distortion to the electron cloud (and thus the change in polarizability) is also predominantly in the same plane as this is where the atomic motion alters the bond lengths. This is schematically shown in Figure 5d. Since the hyperpolarizability is given by the product of these two factors (see Figure 5d), it must be dominated by elements contained within this same plane.(*92*, *97*) With this in mind, the observation of substantial contributions of $CH_2$ in $\chi^{(2)}_{yyx}(f_1)$ and $\chi^{(2)}_{xxx}(f_1)$ with simultaneously no contribution in $\chi^{(2)}_{zzx}(f_1)$ means that this plane of the contributing $CH_2$ groups must lie entirely in the x-y plane and thus parallel to the surface (i.e. with a tilt angle of $\theta \approx 90°$ and a twist angle of $\psi \approx 0°$). Any tilt of the $CH_2$ plane from either of these orientational parameters would lead to increasing contributions in one or more of the other susceptibility elements. Such an orientational picture with essentially in-plane $CH_2$ groups is consistent with the alkyl tails being almost perfectly normal to the surface (shown in Figure 5e). Given that the well-packed monolayer, especially in the LC phase, is expected to have almost 'upright' tails, this result aligns well with the expected conformation.

This finding is very interesting since the presence of such $CH_2$ signals in SFG spectra is typically associated with conformational (gauche) "defects" (randomly distributed kinks in the aliphatic chains). The extracted orientational property of the observed $CH_2$ mode with its in-plane nature clearly does not match such an interpretation but shows that the signal arises from a well-ordered in-plane packing structure. The trouble with this picture, however, is that, for such up-right packing, the tails are generally expected to adopt the 'all-trans' conformation and thus yield an equal number of oppositely oriented $CH_2$ groups, which should result in a vanishing SFG signal. Possible scenarios that would explain the emergence of this signal are that the different $CH_2$ groups are in slightly different local environments and thus yield frequency-shifted resonances that do not entirely cancel or the presence of stress induced deviations in the aliphatic chains from an 'all-trans' conformation. In the first case one would expect to see two close, but distinct contributions with opposite signs in the spectrum, which is not observed. In the second scenario, the $CH_2$ signals (the symmetric stretch vibrations) along the aliphatic chain would not perfectly cancel due to a chain twist, giving rise to a single residual $CH_2$ mode that resembles a $CH_2$ group that is parallel to the sample surface, in agreement with the obtained result. A final confirmation that the $CH_2$ signals indeed originate from such a twisted conformation of the aliphatic chains can be obtained by analyzing the individual in-plane rotational phases of $CH_2$ and $CH_3$ modes which, in case of a twisted chain, should show a phase shift of 90°. For such an analysis, however, it needs to be considered that DPPC contains two aliphatic chains that do not necessarily point in the same in-plane direction. In fact, a previous publication shows that the two $CH_3$ groups indeed point into distinctly different x-y directions with a separation angle of 150°.(*118*) In consequence, the experimentally observed rotational phase only corresponds to the average direction of the two chains making the direct comparison of the $CH_2$ and $CH_3$ orientations difficult. For a rigorous analysis, a separation of the vibrational signals from the two chains is required e.g. by partial deuteration of the DPPC molecule. Such an extended study is currently planned.

***Quantifying the Tilt Angle and In-Plane Packing Structure***

The discussion in the previous section demonstrates the deep insight into the molecular structure that can be gained from rotational SFG microscopy measurements by performing a simple qualitative analysis of the different nonlinear responses. To go a step further, we now turn to a fully quantitative data analysis for the determination of important parameters such as the tilt angles and degree of packing order. For simplicity, we use the $CH_3$ symmetric stretch for this analysis as it is independent of the molecular twist angle, which highly reduces the complexity of the mathematical description. The expressions for the different tensor elements can then be written as in Eqs. 10-14,

$$\chi^{(2)}_{xxx}(f_1) = \frac{1}{4}\rho\beta_{ccc}\langle\sin\theta\,[R(1+3\cos^2\theta)+3\sin^2\theta]e^{-i\phi}\rangle \quad (10)$$

$$\chi^{(2)}_{yyx}(f_1) = \frac{1}{4}\rho\beta_{ccc}\langle\sin\theta\,[R(3+\cos^2\theta)+\sin^2\theta]e^{-i\phi}\rangle \quad (11)$$

$$\chi^{(2)}_{xxz}(f_0) = \frac{1}{2}\rho\beta_{ccc}\langle\cos\theta\,[R(1+\cos^2\theta)+\sin^2\theta]\rangle \quad (12)$$

$$\chi^{(2)}_{zzx}(f_1) = \rho\beta_{ccc}\langle\sin\theta\,[R\sin^2\theta+\cos^2\theta]e^{-i\phi}\rangle \quad (13)$$

$$\chi^{(2)}_{zzz}(f_0) = \rho\beta_{ccc}\langle\cos\theta\,[R\sin^2\theta+\cos^2\theta]\rangle \quad (14)$$

where $R = \frac{\beta_{aac}}{\beta_{ccc}}$ is the hyperpolarizability ratio that has an established literature value of 2.2 for aliphatic methyl groups.(*97*)

In order to simplify the orientational averaging present in Eqs. 10-14, will assume the tilt angle to have a narrow distribution locally (at the single pixel level), such that it can be effectively described by a $\delta$-function. While a perfect $\delta$-distribution is clearly not a precise representation of the real distribution, we note that the strong presence of $CH_3$ SFG signals indicates a general close-packing of the lipids in both phases such that their tail-groups are largely upright and thus should display a high degree of out-of-plane packing order. This picture is thus consistent with a rather narrow distribution of tilt angles when integrating over the sub-micron level. It is therefore reasonable to assume that taking this distribution to be a $\delta$-function does not yield significant misconceptions about the lipid structure. We therefore rewrite Eqs. 10-14 as in Eqs. 15-19,

$$\chi_{xxx}^{(2)}(f_1) = \frac{1}{4}\rho\beta_{ccc}\sin\theta\,[R(1+3\cos^2\theta)+3\sin^2\theta]\langle e^{-i\phi}\rangle \tag{15}$$

$$\chi_{yyx}^{(2)}(f_1) = \frac{1}{4}\rho\beta_{ccc}\sin\theta\,[R(3+\cos^2\theta)+\sin^2\theta]\langle e^{-i\phi}\rangle \tag{16}$$

$$\chi_{xxz}^{(2)}(f_0) = \frac{1}{2}\rho\beta_{ccc}\cos\theta\,[R(1+\cos^2\theta)+\sin^2\theta] \tag{17}$$

$$\chi_{zzx}^{(2)}(f_1) = \rho\beta_{ccc}\sin\theta\,[R\sin^2\theta+\cos^2\theta]\langle e^{-i\phi}\rangle \tag{18}$$

$$\chi_{zzz}^{(2)}(f_0) = \rho\beta_{ccc}\cos\theta\,[R\sin^2\theta+\cos^2\theta] \tag{19}$$

where the orientational averaging is now restricted to any dependence on the in-plane azimuth angle, $\phi$.

On inspection of Eqs. 15-19, it is clear that there are four distinct ways of extracting the tilt angle without any assumptions on the molecular density or in-plane distribution (or even the value of the $\beta_{ccc}$ hyperpolarizability component), namely using the ratio of any pair of $\chi_{xxx}^{(2)}(f_1)$, $\chi_{yyx}^{(2)}(f_1)$, and $\chi_{zzx}^{(2)}(f_1)$, or the ratio of $\chi_{xxz}^{(2)}(f_0)$ to $\chi_{zzz}^{(2)}(f_0)$. While these, in theory, should be equivalent methods, we opt for the latter ratio as it combines data with higher signal-to-noise and $CH_3$ resonances that are more easily isolated from their neighboring $CH_2$ contributions in the spectrum. An expression for this ratio is given in Eq. 20, which is then solved for the tilt angle in Eq. 21.

$$\frac{\chi_{xxz}^{(2)}(f_0)}{\chi_{zzz}^{(2)}(f_0)} = \frac{[R(1+\cos^2\theta)+\sin^2\theta]}{2[R\sin^2\theta+\cos^2\theta]} \tag{20}$$

$$\theta = \sin^{-1}\sqrt{\frac{2\left(R-\frac{\chi_{xxz}^{(2)}(f_0)}{\chi_{zzz}^{(2)}(f_0)}\right)}{(R-1)\left(1+2\frac{\chi_{xxz}^{(2)}(f_0)}{\chi_{zzz}^{(2)}(f_0)}\right)}} \tag{21}$$

Figure 6b then shows a map of the calculated $CH_3$ tilt angle, where the contrast between the LC and LE phases becomes immediately clear, with the former showing tilt angles of ~30º and the latter ~60º. Given the known tetrahedral bond arrangement in the aliphatic chain, a completely 'upright' chain would give rise to a $CH_3$ tilt angle of ~35º. It is thus clear that the molecular tilt of DPPC in the LC domains is incredibly small, while the LE phase has a more pronounced tilt (average molecular tilt of roughly 25°). Again, this aligns well with the expected nature of the two phases being 'condensed' and 'expanded', respectively, with the latter offering more space for the molecules to tilt over. Equally, the essentially upright tails also agree with the earlier conclusions of the $CH_2$ groups being in-plane. A simple schematic of this difference is also shown below the tilt angle map in Figure 6b.

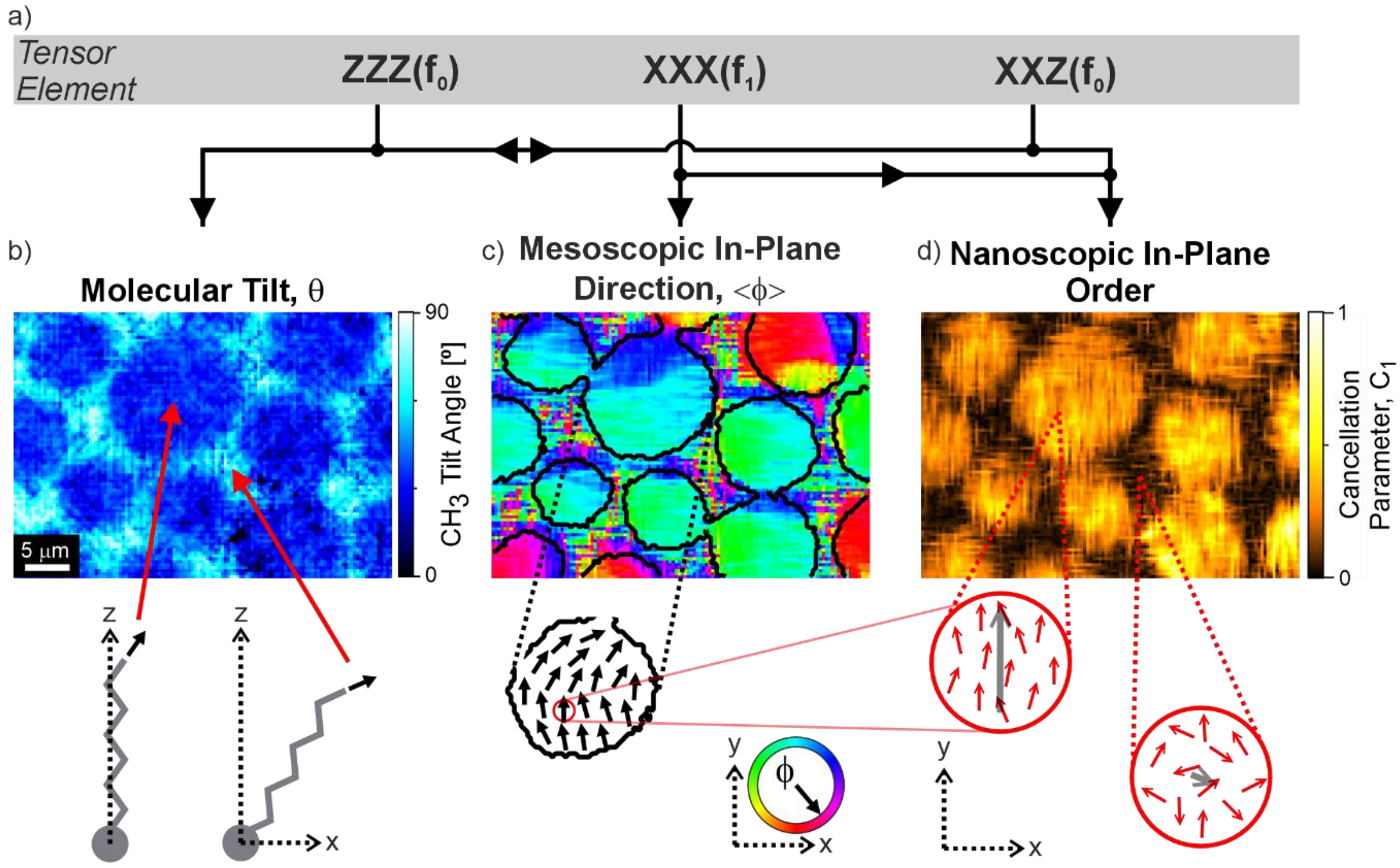

*Figure 6:* ***Extracted tilt angle and in-plane structure.*** *(a) Schematic workflow showing how the different tensor elements are combined to extract structural parameters. (b) Image of the $CH_3$ tilt angle, calculated from the $\chi^{(2)}_{xxz}(f_0)$ and $\chi^{(2)}_{zzz}(f_0)$ tensor elements. Also shown is a simple schematic highlighting the different molecular tilt conformations between the LC and LE phases. (c) Image of the in-plane rotational phase, representing the average molecular direction in terms of its azimuth, $\langle\phi\rangle$. This molecular directionality is highlighted for an example LC domain, showing the curved in-plane packing structure at the mesoscopic scale. (d) Image of the $C_1$ cancellation parameter, calculated using the ratio of the measured $\chi^{(2)}_{xxx}(f_1)$ data and that predicted for a perfectly ordered system using the $\chi^{(2)}_{xxz}(f_0)$ response and calculated tilt angle (therefore combining all three tensor elements). Simple schematics of this ordering are shown for the LC and LE phases, representing the in-plane orientational distribution within a single pixel, i.e. at the nanoscopic scale – the red arrows represent individual molecules with the gray arrows indicating the amount of order and net direction within that pixel.*

If we turn back to the expressions for the $\chi^{(2)}_{xxx}(f_1)$, $\chi^{(2)}_{yyx}(f_1)$, and $\chi^{(2)}_{zzx}(f_1)$ elements given in Eqs. 15, 16, and 18, we showed that they all maintain a dependence on $\phi$, with its distribution still being subjected to the orientational averaging. As with $\theta$, a $\delta$-function for the $\phi$-distribution would clearly remove this averaging from the expression. Unlike the tilt angle, however, such a narrow distribution in the in-plane orientation is far from realistic, and generally a poor assumption. As discussed above, the $\chi^{(2)}_{xxx}(f_1)$, $\chi^{(2)}_{yyx}(f_1)$, and $\chi^{(2)}_{zzx}(f_1)$ components are distinctly lower in amplitude than $\chi^{(2)}_{xxz}(f_0)$ and $\chi^{(2)}_{zzz}(f_0)$, especially for the LE where they essentially show no significant signals, indicating at least a moderate degree of in-plane disorder. As such, we rather rewrite the orientational averaging over this rotational phase in terms of an average phase (direction) and a cancellation parameter, $C_1$, which represents the relative amount of signal reduction caused by the in-plane orientational distribution, i.e. a value of 1 indicates no cancellation (perfect order) and 0 complete cancellation (isotropic distribution). This is shown for the 1-fold $\chi^{(2)}_{xxx}(f_1)$ contribution in Eq. 22.

$$\chi^{(2)}_{xxx}(f_1) = \frac{1}{4}\rho\beta_{ccc}\sin\theta\,[R(1+3\cos^2\theta)+3\sin^2\theta]C_1 e^{-i\langle\phi\rangle} \tag{22}$$

Isolating the rotational phase of the $\chi^{(2)}_{xxx}(f_1)$ component then yields the average azimuth, which is shown in Figure 6c. Identical maps of this in-plane direction can also be obtained through the other components ($\chi^{(2)}_{yyx}(f_1)$ or $\chi^{(2)}_{zzx}(f_1)$), or even simply from the 1-fold components of the effective susceptibilities in the different polarization combinations, as shown in Supplementary Information. On closer inspection of the average directions, it is clear that the LC domains display long-range order, but nevertheless have a slowly changing molecular direction akin to a curved packing structure. This aligns with our previous observations of this in-plane packing in these systems(*113*), where the domains were shown to adopt a spiral packing motif that stems from the molecular chirality of DPPC.

A simple rearrangement of Eq. 22 then yields the expression for the cancellation parameter $C_1$ in Eq. 23 in terms of the measured $\chi^{(2)}_{xxx}(f_1)$ magnitude. In addition to $\theta$, however, determining $C_1$ from this expression clearly also requires knowledge of the density and hyperpolarizability, which are not generally known. This dependence, however, can be circumvented by also using the $\chi^{(2)}_{xxz}(f_0)$ (or $\chi^{(2)}_{zzz}(f_0)$) element, as shown in Eq. 24, resulting in an expression that makes use of only measured data and the already extracted tilt angle.

$$C_1 = \frac{\left|\chi^{(2)}_{xxx}(f_1)\right|}{\frac{1}{4}\rho\beta_{ccc}\sin\theta\,[R(1+3\cos^2\theta)+3\sin^2\theta]} \tag{23}$$

$$= \frac{\left|\chi^{(2)}_{xxx}(f_1)\right|}{\chi^{(2)}_{xxz}(f_0)}\frac{2[R(1+\cos^2\theta)+\sin^2\theta]}{\tan\theta\,[R(1+3\cos^2\theta)+3\sin^2\theta]} \tag{24}$$

Figure 6d shows a map of this extracted cancellation parameter $C_1$. As with the tilt angle, this also shows clear structural contrast between the LC and LE phases. Specifically, the LC domains present values of ~0.3-0.5, while the LE phase essentially has complete cancellation ($C_1 \approx 0$). This confirms that the LE phase is essentially isotropic in-plane at the single pixel level leading to a complete cancellation of the in-plane signals. Conversely, the LC phase does show considerable in-plane order, but nevertheless displays a remarkable large amount of in-plane cancellation. This hence categorically demonstrates that the $\phi$-distribution is far from being both isotropic or a $\delta$-function in the LC domains. This result is surprising as a relatively large amount of nanoscopic in-plane disorder seems contradictory to the observation of a well-defined long-range curvature in the average x-y direction. As shown in a recent publication, more insight into the shape of the $\phi$-distribution function on a single pixel level can be obtained by including the higher order rotational harmonics in the analysis. The results demonstrate that $\phi$ in fact follows a bimodal distribution with a separation of approximately 150 deg. which arises from each molecule having two aliphatic chains with distinctly different x-y orientations.(*118*) It is thus the bimodal nature of the $\phi$-distribution function and not molecular disorder that is the dominant source for the observed signal reduction in the LC domains which resolves the apparent contradiction.

***Determining Density and Composition***

Having extracted information on the tilt angle and in-plane distribution, the only remaining structural parameter in Eqs. 15-19 is the molecular density, $\rho$. Importantly, however, this is only the molecular density of the molecule being probed—in this case DPPC, given that we are analyzing the C-H stretches. In order to get information on the true density differences between the LC and LE phases, as well as any differences in composition, we thus require similar information from the other lipid, dPOPC. For this, we therefore also recorded hyperspectral SFG imaging with azimuthal-scanning in the C-D stretching region. However, unlike for the C-H region, a full decomposition into different tensor elements is in this case not feasible due to the significantly weaker signals in the C-D region. There are three reasons for this: Firstly, the C-D transition dipole moment is notably ~30-40% weaker than for C-H. Secondly, the dPOPC only makes up 20% of the film, and thus is four times less abundant than DPPC on average. Finally, as POPC is unsaturated, it generally forms a less well-packed film and thus offers weaker SFG signals. Nevertheless, taking only the 0-fold PPP contribution is sufficient to extract the density in this case.

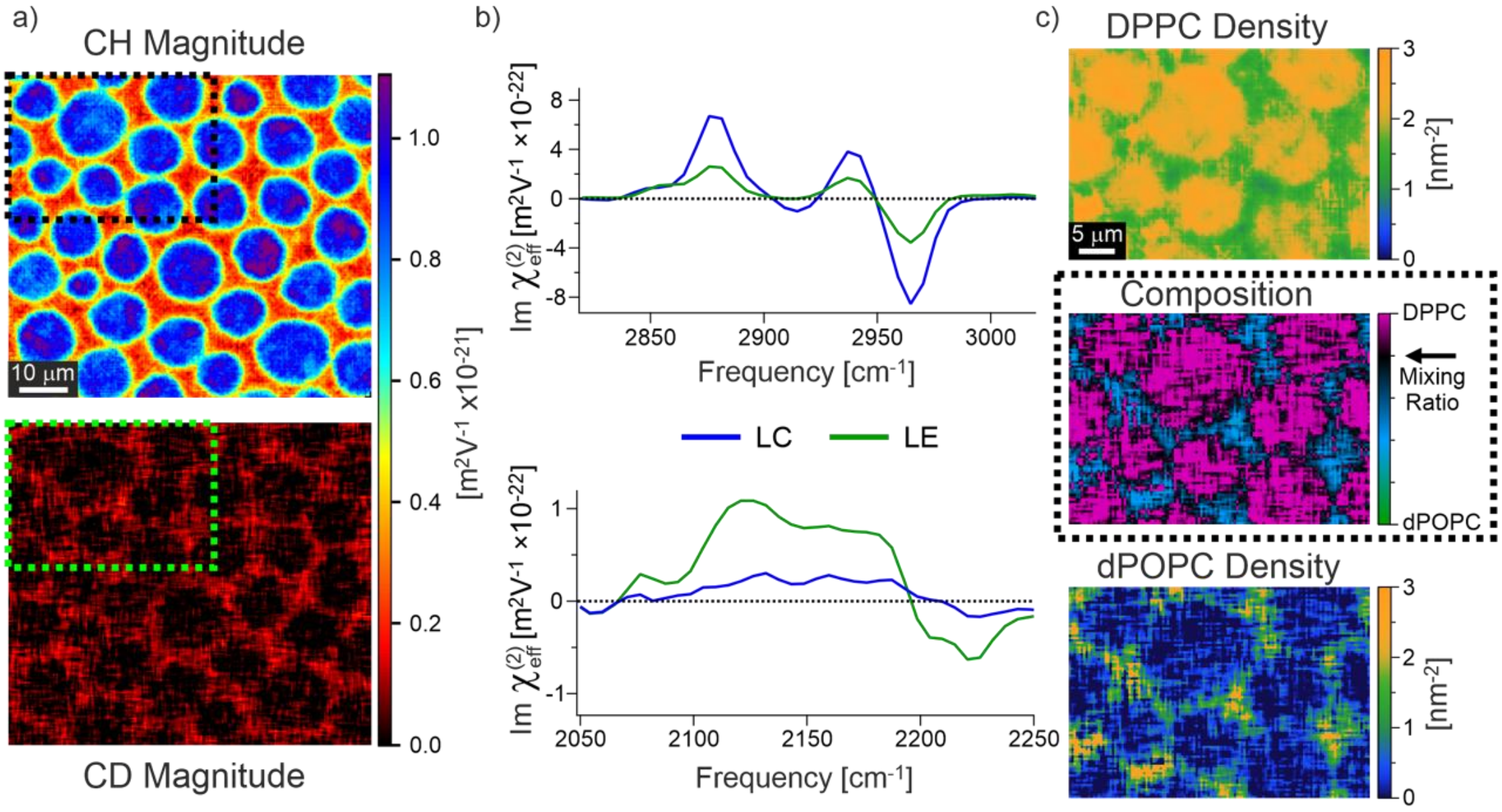

*Figure 7:* ***Density and composition analysis.*** *(a) Magnitude images of the effective PPP response (including Fresnel factors and beam geometry factors) in the 0-fold contribution for both C-H and C-D, given as absolute values. (b) Spectra (imaginary parts) averaged over each of the two structural phases, LC (blue) and LE (green), also shown in absolute units. (c) Extracted images of the molecular density for DPPC and dPOPC, using the PPP data and calculated tilt angle from Figure 6b, and the known average density and mixing ratio of the lipids. The presented surface region corresponds to that*

*indicated by the dashed boxes in (a). Also shown is a composition image calculated from the two density maps, where black is used to represent the 4:1 mixing ratio.*

Figure 7a shows 0-fold magnitude images for the PPP polarization combination (still including the experimental factors) in the C-H and C-D stretching regions, both presented in absolute units and on the same scale. The generally lower signal expected from dPOPC becomes immediately evident on comparison of the two images, but nevertheless still shows a defined contrast between the two phases. Interestingly, while the LC domains show a much larger response in the C-H region, as has been the case throughout this study, the C-D region shows the opposite—almost no signal within the LC domains and a pronounced response from the LE phase. This difference is emphasized in the spectra shown in Figure 7b, which have been averaged over each phase. While the amplitudes of the spectra clearly highlight this difference, it is important to note the lack of 1-to-1 comparability between the C-H and C-D line-shapes, which stems from the fundamental differences between the absolute and relative vibrational frequencies of $CD_2$ and $CD_3$ compared to their protiated counterparts.(*131*)

At this point, we also note that the CD magnitudes, maximizing at $\sim 1 \times 10^{-22}$ $m^2V^{-1}$, highlight the true sensitivity of the SFG microscope. Specifically, it shows that clear image contrast can be obtained from signals even down to a ~1/10$^{th}$ of a monolayer. In fact, this amplitude of the effective susceptibility is similar to the O-H stretching response from the pure air-water interface, which is renowned in SFG spectroscopy to yield an extraordinarily weak signal owing to its broad orientational distribution.(*66*, *86*) This sensitivity is even more remarkable when considering these signals are from C-H and C-D resonances, and not from stronger SFG-active chromophores.

Using the magnitude maps presented in Figure 7a, along with the extracted tilt angle map from Figure 6b, Figure 7c then shows the retrieved density maps of the two lipids (for details about the calculation of these maps see Supplementary Information). These have also been converted into absolute units using the known average density of the film, which was determined from the Langmuir isotherm data obtained when compressing and casting the film. It is, however, important to note that we have assumed the dPOPC to locally adopt the same tilt angle as DPPC (in both phases). This assumption was required as its tilt angle cannot be independently extracted due to the absence of the required tensor elements for the CD region. While this assumption may not be entirely accurate, it is based on the logical conclusion that the tilt angle of a molecule will be predominantly dictated by the conformation of their nearest neighbors, with similar tilt angles generally leading to better packing and stronger intermolecular interactions.

Overall, these density maps confirm the observations from the magnitude images in Figure 7a that the DPPC is primarily concentrated in the LC phase, with the dPOPC almost exclusively being in the LE phase. This compositional difference between the two phases is also made clear in the center of Figure 7c, which shows the LC phases to consist almost entirely of DPPC and the LE phase to have a more even mixture. Such compositional enrichment from the average mixing ratio of 4:1 (represented as black in the composition map in Figure 7c) aligns with suggestions from previous fluorescence and Raman microscopy investigations(*39*, *40*, *132*) and generally highlights the molecular specificity of the intermolecular interactions, with the optimum packing arrangement arising from homo-molecular assemblies. The resulting compositional heterogeneity in such films clearly represents a fine balance between the enthalpic difference of homo- and hetero-molecular interactions and the entropic drive towards a homogeneous film.

While the discussion above highlights the deep structural insight into these phospholipid monolayers that has been obtained in this study, it is important to mention that the presented analysis only exploits a subsection of the recorded data. A rigorous determination of structural parameters such as the Euler angles for every detected pixel has only been performed on the symmetric stretching mode of the $CH_3$ group. The same type of analysis could obviously be performed on all remaining vibrational modes in the spectrum and would consequently yield individual Euler angles for each of them. Furthermore, other functional groups contained in the phospholipid molecules such as the carbonyl groups could additionally be probed by tuning to the respective frequency range. Since all these modes are located on the same molecule the knowledge of all individual Euler angles can in principle be used to determine the exact molecular conformation, an approach that was partly demonstrated by the comparison of the orientational properties of $CH_3$ and $CH_2$ modes. Such a global data analysis would obviously require a more sophisticated and complex analysis approach (e.g. specialized algorithms), however, this perspective along with the enormous potential of such approach defines a clear future goal of our technical development.

## Conclusions

As shown above, rotational SFG microscopy can provide fascinating insight into the heterogeneous molecular structure at interfaces. Our study on monolayers of mixed phospholipids (POPC and DPPC) reveals their packing structure in unprecedented detail: Upon compression the mixed layers form distinct condensed domains (LC) of about 10 µm in diameter that are surrounded by a liquid expanded (LE) phase. While the overall molecular density in both phases is surprisingly similar (ca. 3 molecules per $nm^2$) the two phases largely deviate in composition and molecular packing. While the LC domains almost exclusively consist of DPPC, the LE phase shows a mixture between DPPC and POPC lipids. In the LC domains, the DPPC molecules are densely packed in an almost crystalline form, which includes a high degree of molecular order parallel and perpendicular to the surface. The molecular orientation in the surface plane is thereby not constant across the domains but shows a pronounced long-range curvature. This curvature is part of a macroscopic spiraling in-plane packing structure which is induced by the molecular chirality of the DPPC molecule. In this packing structure the aliphatic chains point almost perfectly up (orthogonal to the sample surface) and seem to possess a twisted chain conformation while the individual in-plane orientation of the two chains is separated by approximately 150 deg. In stark contrast, the DPPC in the LE phase shows no molecular order parallel to the surface and much larger molecular tilt angles with an average of ca. 25 deg. This increased average molecular tilt is clear indication that the molecular order in the LE phase is also greatly reduced in the out-of-plane direction (orthogonal to the surface).

This insight that is obtained in such experiments can be divided into two parts: i) information about the local (single pixel) molecular-level structure within the resolution limit of currently about 900 nm that is of statistical nature and ii) information on the microscale heterogeneity of these structural parameters obtained by the wide-field imaging. That way the structural insight gained by such experiments span large length-scales, from the molecular up to the macroscopic level.

More specifically, from each pixel inside the acquired images, we obtain information on average molecular orientations (tilt and azimuth) including the width (orientational order) and functional form of the orientational distribution. Moreover, the analysis of these parameters for the vibrations originating from different parts of the molecules allows for determining specific molecular conformations and, as demonstrated in this study, such measurements allow for determining molecular compositions and overall density. Finally, the presented analysis can in principle be extended to also obtain information on the intermolecular environment and interactions by analyzing vibrational line-shapes and shifts in resonance frequencies. The spatial variation of all these parameters within the acquired wide-field image then reveals mesoscopic molecular packing structures (such as the observed curved in-plane packing of DPPC shown here) and general heterogeneity in the molecular layer (e.g. phase separated domains). Importantly, all this can be done in a label-free manner, under ambient conditions, and even for the case of buried interfaces.

However, aside from these wide-ranging capabilities of the presented microscopy technique, it is also important to point out its limitations. One important aspect is that measurements as those presented in this study are very time consuming. Although the intrinsic time resolution of the microscopy is only limited by the decay of the vibrational coherence (typically on the order of picoseconds), performing a complete series of experiments including high frequency resolution, sample rotation, and measurements at different polarization combinations obviously come with relatively long acquisition times (currently on the order of several hours). Such experiments therefore require sample systems with high temporal stability which makes studies of dynamic structural changes rather difficult unless these changes are induced by some external trigger and, for studying fast dynamics, are highly repeatable.

A second limitation is that SFG can only probe the anisotropic part of a structure. While anisotropy is a common phenomenon at interfaces and often contains the most relevant information, there are also sample classes and research questions that specifically require knowledge of the isotropic structural components. A prominent example related to the present study is the out-of-plane structure in symmetric lipid bilayers. Because of the opposite orientation of the lipids along the surface normal, a large fraction of the SFG signals would cancel, which reduces the structural insight that can be gained. For that reason, it seems very appealing to combine rotational SFG microscopy with e.g. coherent Raman microscopy to overcome this limitation in future studies.

Nevertheless, the structural insight provided by rotational SFG microscopy clearly goes beyond the insight that can be obtained by other, more established experimental tools and allows for a completely new type of experimental studies. Examples include the investigation of precise structure-function relationships within biophysical systems, or the important connection between molecular and structural chirality, and how this can be imprinted on the surrounding media. Furthermore, it is important to note that the applicability of the technique is not restricted to molecular interfaces. As shown in recent work, exactly the same measurement concept can be used to obtain important insight into the structures of stacked 2D materials(*116*) as well as to study ordering effects of magnetic domains in antiferromagnets(*117*). With its unique capability and broad applicability, the presented microscopy technique fills an important experimental gap for future studies and is anticipated to become a valuable addition to the toolbox of characterization techniques of interfacial structures.

## Experimental

### *Sample Preparation*

The phospholipids 1,2-dipalmitoyl-sn-glycero-3-phosphocholine (DPPC, >99%) and $d_{82}$-1-palmitoyl-2-oleoyl-glycero-3-phosphocholine (dPOPC, >99%) were obtained from Avanti Polar Lipids (Alabaster, AL, USA) and dissolved in chloroform (99.0–99.4%, VWR International GmbH, Darmstadt, Germany) to prepare stock solutions with concentrations of 1 mg $mL^{-1}$. Mixed lipid monolayers were prepared by pre-mixing DPPC and dPOPC in a 4:1 mass ratio prior to spreading the solution dropwise onto the surface of a PTFE Langmuir–Blodgett trough (MicroTrough G1, Kibron, Helsinki, Finland) filled with ultrapure water (Milli-Q, 18.2 MΩ cm, <3 ppb total organic carbon). Prior to the drop casting, the trough was thoroughly cleaned with chloroform and ultrapure water, and the water surface was repeatedly aspirated until the measured surface pressure remained within 0.1 mN $m^{-1}$at full compression. The platinum Wilhelmy pressure sensor was additionally flamed and rinsed with both chloroform and ultrapure water to minimize contamination. After deposition, the monolayers were allowed to equilibrate for approximately 5 min to ensure complete chloroform evaporation. The films were subsequently compressed to a surface pressure of 20 mN $m^{-1}$at a barrier speed of 10 mm $min^{-1}$, followed by a further equilibration period of 2 h during which oxidation due to ambient ozone exposure could occur.

For microscopy measurements, the monolayers were immobilized to suppress intrinsic lateral dynamics as well as Bénard–Marangoni convection induced by local heating of the aqueous subphase(*133*). Transfer of the monolayers onto solid substrates was achieved by Langmuir–Blodgett deposition using a LayerX dipper (Kibron, Helsinki, Finland) operated at a dipping speed of 2 mm $min^{-1}$. Ultra-flat fused silica windows (Korth Kristalle, Altenholz, Germany; 5 mm thickness, 25.4 mm diameter, surface roughness <2 nm) served as substrates. Prior to deposition, the substrates were cleaned with chloroform and ultrapure water and subsequently treated with UV–ozone for at least 30 min using a UV/Ozone ProCleaner Plus system (BioForce Nanosciences, Virginia Beach, VA, USA).

### *SFG Microscopy Measurements*

Broadband infrared and visible excitation pulses were generated by optical parametric amplification of the 800 nm output from an amplified Ti:sapphire laser system (7W, 1kHz, 30 fs pulse length, Coherent, USA). The sample was irradiated at an incidence angle of 36° through a custom hole in a reflective objective (0.78 NA, Pike Technologies, Madison, WI, USA) where the beams illuminate a sample area of roughly 150 µm in diameter. Following interaction with the sample, the generated SFG signal was spectrally filtered, temporally overlapped, and imaged onto a thermoelectrically cooled CCD camera (ProEM-HS:1024BX3, Teledyne Princeton Instruments, Trenton, NJ, USA).

Hyperspectral SFG image stacks were recorded over temporal delays ranging from −300 to 3000 fs in 2 fs increments. Phase and amplitude referencing was performed using equivalent measurements on z-cut quartz acquired over a delay range from −300 to 300 fs. The datasets presented correspond to the co-average of multiple measurements, with a minimum of four scans acquired for each azimuthal orientation. Azimuthal scans were performed by rotating the sample about the surface normal in discrete angular increments (90° steps for the SSS and CD data, 45° steps for all other measurements) while maintaining a fixed optical alignment. Polarization-dependent measurements were carried out by varying the polarization combinations of the incident infrared and visible excitation beams together with the polarization of the local oscillator (LO) reference beam. For each polarization combination a separate reference measurement on z-cut quartz was performed. Measurements in different infrared spectral regions were achieved by tuning the center frequency of the infrared output of the optical parametric amplifier.

The raw hyperspectral images were processed by subtracting detector dark counts and removing residual non-resonant background contributions. The processed datasets were numerically back-rotated and spatially aligned, generating four-dimensional datasets containing both temporal and rotational information. Fourier transformation along the temporal axis yielded vibrational spectra, whereas Fourier transformation along the azimuthal dimension provided rotational frequencies in form of real and imaginary parts (x and y projections) or magnitude and phase (x-y angle) as functions of vibrational frequency.

***Data Treatment***

The determination of the hyperspectral images of the individual tensor components (Figure 5) was performed using the raw data from the rotational measurements at the different polarization combinations (PPP, SSP, SSS) in combination with Eqs. 5-9. For better visualization of the 3D hyperspectral datasets in Figures 4, 5, and 7, a singular value decomposition (SVD) was applied to reduce the dimensionality of the hyperspectral images by converting the spectral frequency axis into a set of principal component spectra. It was found that for each data set one single dominant SVD component sufficiently reproduces the measured data. The results in Figures 4 and 5 are therefore presented by i) the respective dominant SVD spectrum (spectra are normalized to the largest peak) and ii) a 2D image representing the amplitude of this spectrum for each pixel (sample position). For the 1-fold rotational components a combined SVD analysis of the real and imaginary parts of the rotational Fourier transformation was performed yielding, besides the SVD spectrum, two separate amplitude images. The presented single 1-fold 2D images shown in Figures 4 and 5 then represent the absolute value (magnitude) images calculated from the two respective amplitude images (see supplementary information).

Spectral analysis of the hyperspectral datasets in Figure 5 was done by performing a global multi-Lorentzian fit (IgorPro, WaveMetrics) on real and imaginary parts of all five presented SVD spectra. The resulting fit parameters are shown in the supplementary information. The tilt angle map presented in Figure 6b was obtained by scaling the 2D SVD images in Figure 5 by the corresponding amplitude of the $CH_3$ symmetric stretch signals (obtained by the global fit of the SVD spectra) and subsequent application of Eq. 21. The map of molecular in-plane directions in Figure 6c was obtained by extracting the in-plane phase map from the two individual SVD amplitude images (real and imaginary parts of the rotational Fourier transformation). Finally, the in-plane order map in Figure 6d is determined by applying Eq. 24 to the SVD images presented in Figure 5 (scaled by the amplitude respective $CH_3$ symmetric stretch signals in the corresponding SVD spectra) including the obtained 2D map of tilt angles shown in Figure 6b. The data presented in Figure 7 correspond to the respective SVD images (7a) along with the average spectra (7b) of the LC and LE phases obtained from the corresponding raw hyperspectral images. For the calculation of the densities and compositions the SVD images (Figure 7a) are used in combination with the tilt angle map (Figure 6b) and additional information on average densities from the Langmuir isotherms (for more details see supplementary information).

# Visualizing the Hidden Architecture of Molecular Films with Phase-Resolved Rotational SFG Microscopy

## Supplementary Information

Ben John, Tuhin Khan, Nasim Mirzajani, Martin Wolf, Alexander P. Fellows*, and Martin Thämer*

Fritz-Haber-Institut der Max-Planck-Gesellschaft, Faradayweg 4-6, 14195, Berlin, Germany

*Corresponding Authors

Email: fellows@fhi-berlin.mpg.de

Tel: +49 (0)30 8413 5140

Email: thaemer@fhi-berlin.mpg.de

Tel: +49 (0)30 8413 5145

This supplementary information provides additional details on experimental parameters and analytical procedures employed to obtain the resulting data presented in the main text. First, the calculation of the number of molecules that give rise to the signal detected on a single pixel is presented followed by a detailed description of the normalization procedure to yield the resulting data in absolute units. This includes a table with all relevant experimental parameters used within this procedure. In a subsequent section, details on the SVD analysis of the recorded hyperspectral images are presented, which is followed by a comparison of the in-plane phase images obtained from the measurements at different polarization combinations. Finally, a list of fit parameters from the global fit of the different SFG spectra is shown followed by the presentation of the procedure used for calculating the molecular densities and compositions.

### *Number of Molecules per Imaged Pixel*

The theoretical sample area that corresponds to one pixel is 0.1 $\mu m^2$ (pixel size on chip is 169 $\mu m^2$ and the magnification of the objective is 40x). Although the resolution of the SFG microscope is only just below 1$\mu$m, the light level detected on a single pixel correspond to the light level emitted by the number of molecules that are contained in this theoretical sample spot of 0.1 $\mu m^2$. Given the determined molecular density of DPPC inside the LC domains of ca. 2 $nm^{-2}$ and the fact that there are two $CH_3$ groups per DPPC molecule we obtain a number of ca. $2 \cdot 10^5$ molecules or $4 \cdot 10^5$ $CH_3$ groups that are present within this theoretical spot size.

### *Conversion of SFG Microscopy Data into Absolute Units*

The presented results are given in absolute units of the susceptibility, which is obtained by normalization of the acquired data to a separate reference measurements of a z-cut alpha quartz crystal. However, the normalization cannot be done by simply dividing the sample signal by the reference signal since the correct conversion includes several experimental factors that need to be properly accounted for. The measured heterodyned signal $S$ is generally given by Eq. S1,

$$S \propto E_{SFG} E_{LO}^{*} + CC \quad (S1)$$

where $E_{SFG}$ is the generated SFG field in the reflection direction and $E_{LO}^{*}$ is the complex conjugate of the local oscillator (LO) field reflected by the sample surface. $E_{SFG}$ depends on incidence angle ($\theta$), polarization combination, nonlinear Fresnel factors ($L$), the pump fields $E(\omega_1)$ and $E(\omega_2)$ as well as the evolution of the nonlinear susceptibility with sample depth. $E_{LO}^{*}$ meanwhile depends on the linear reflection coefficient $R_{S/P}^{LO}$ and thus on the refractive index of the substrate material, the incidence angle, and the polarization of the LO (note, in the balanced imaging scheme employed in our microscopy approach the LO polarization is always orthogonal to the polarization of the SFG field). Some of these quantities change when replacing the sample (molecular monolayer on fused silica) with the reference quartz crystal, therefore requiring a detailed mathematical description of the measured signal sizes. For the reference quartz measurement (superscript Q) the measured signal can be expressed as in Eqs. S2-S4,(*66*, *134*)

$$S_{SSP}^{Q} \propto \cos\theta \cdot L_y^Q(\omega_3) L_y^Q(\omega_2) L_x^Q(\omega_1) \frac{i}{\Delta k_z^Q} \chi_{y'y'x'}^{(2),Q} \sqrt{R_P^{LO,Q}} \left(E(\omega_2)E(\omega_1)E_{LO}^{*} + CC\right) \quad (S2)$$

$$S_{PPP}^{Q} \propto \cos^3\theta \cdot L_x^Q(\omega_3) L_x^Q(\omega_2) L_x^Q(\omega_1) \frac{i}{\Delta k_z^Q} \chi_{x'x'x'}^{(2),Q} \sqrt{R_S^{LO,Q}} \left(E(\omega_2)E(\omega_1)E_{LO}^{*} + CC\right) \quad (S3)$$

$$S_{SSS}^{Q} \propto L_y^Q(\omega_3) L_y^Q(\omega_2) L_y^Q(\omega_1) \frac{i}{\Delta k_z^Q} \chi_{x'x'x'}^{(2),Q} \sqrt{R_P^{LO,Q}} \left(E(\omega_2)E(\omega_1)E_{LO}^{*} + CC\right) \quad (S4)$$

where $\Delta k_z$is the z-component of the wavevector mismatch, which is given by $\Delta k_z = k_z^1 + k_z^2 + k_z^3$ for the employed reflection geometry. The tensor elements of the quartz susceptibility are given in the crystal frame (x', y', z') while all other indices are given in the laboratory frame (x, y, z). Because of the z-cut angle z'=z and the relation between x' and x (y' and y) depends on the in-plane orientation of the quartz crystal. For the SSP and PPP measurements quartz is oriented such that x'=x and y'=y whereas for the SSS measurement the crystal must be rotated by 90 degrees such that x'=y and y'=-x (this is necessary since $\chi_{y'y'y'}^{(2),Q}$=0). Note, Fresnel factors for the input ($\omega_1$, $\omega_2$) and output beams ($\omega_3$) generally differ even when neglecting dispersion effects.(*83*) The corresponding sample responses (superscript S) are given by Eqs. S5-S7.

$$S_{SSP}^{S} \propto \chi_{SSP,eff}^{(2),S}(f_n) \sqrt{R_P^{LO,S}} \left(E(\omega_2)E(\omega_1)E_{LO}^{*} + CC\right) \quad (S5)$$

$$S_{PPP}^{S} \propto \chi_{PPP,eff}^{(2),S}(f_n)\sqrt{R_S^{LO,S}}(E(\omega_2)E(\omega_1)E_{LO}^{*} + CC) \qquad (S6)$$

$$S_{SSS}^{S} \propto \chi_{SSS,eff}^{(2),S}(f_n)\sqrt{R_P^{LO,S}}(E(\omega_2)E(\omega_1)E_{LO}^{*} + CC) \qquad (S7)$$

The angular dependence and the Fresnel factors are here included in the effective susceptibilities, as shown in Table 1 in the main text. Combining Eqs. S2-S4 with Eqs. S5-S7 yields Eqs. S8-S10.

$$\chi_{SSP,eff}^{(2),S}(f_n) = \frac{S_{SSP}^{S}}{S_{SSP}^{Q}} \cdot \frac{\sqrt{R_P^{LO,Q}}}{\sqrt{R_P^{LO,S}}} \cos\theta \cdot L_y^Q(\omega_3)L_y^Q(\omega_2)L_x^Q(\omega_1)\frac{i}{\Delta k_z^Q}\chi_{y'y'x'}^{(2),Q} \qquad (S8)$$

$$\chi_{PPP,eff}^{(2),S}(f_n) = \frac{S_{PPP}^{S}}{S_{PPP}^{Q}} \cdot \frac{\sqrt{R_S^{LO,Q}}}{\sqrt{R_S^{LO,S}}} \cos^3\theta \cdot L_x^Q(\omega_3)L_x^Q(\omega_2)L_x^Q(\omega_1)\frac{i}{\Delta k_z^Q}\chi_{x'x'x'}^{(2),Q} \qquad (S9)$$

$$\chi_{SSS,eff}^{(2),S}(f_n) = \frac{S_{SSS}^{S}}{S_{SSS}^{Q}} \cdot \frac{\sqrt{R_P^{LO,Q}}}{\sqrt{R_P^{LO,S}}} L_y^Q(\omega_3)L_y^Q(\omega_2)L_y^Q(\omega_1)\frac{i}{\Delta k_z^Q}\chi_{x'x'x'}^{(2),Q} \qquad (S10)$$

(It is important to note that, while Eqs. S2-S7 are given by proportionalities, they all share the same proportionality constant, which is why Eqs. S8-S10 are given as equalities.)

The relevant experimental parameters for the evaluation of Eqs. S8-S10 are listed in Table S1 below.

| | Fused Silica (FS) | Quartz (Q) |
|---|---|---|
| Incidence angle, $\theta'$ | 36° | |
| $L_x^{IN}$ | 0.876 | 0.852 |
| $L_y^{IN}$ | 0.758 | 0.725 |
| $L_z^{IN}$ | 0.535 | - |
| $L_x^{OUT}$ | 1.124 | 1.148 |
| $L_y^{OUT}$ | 1.242 | 1.275 |
| $L_z^{OUT}$ | 1.842 | - |
| $\chi_{x'x'x'}^{(2),Q} = -\chi_{y'y'x'}^{(2),Q}$ | - | 0.6 pm / V |
| $R_S^{LO}$ | 0.0602 | 0.0766 |
| $R_P^{LO}$ | 0.0160 | 0.0223 |
| $\Delta k_z^{-1}$ | - | 32.14 nm |

*Table S1. Experimental parameters for conversion into absolute units.*

Inserting these values into Eqs. S8-S10 gives conversion constants in Eqs. S11-S13.

$$\chi^{(2),S}_{SSP,eff}(f_n) = \frac{S^S_{SSP}}{S^Q_{SSP}} \times 1.46 \times 10^{-20} m^2 V^{-1} \quad (S11)$$

$$\chi^{(2),S}_{SSS,eff}(f_n) = \frac{S^S_{SSS}}{S^Q_{SSS}} \times 1.53 \times 10^{-20} m^2 V^{-1} \quad (S12)$$

$$\chi^{(2),S}_{PPP,eff}(f_n) = \frac{S^S_{PPP}}{S^Q_{PPP}} \times 9.73 \times 10^{-21} m^2 V^{-1} \quad (S13)$$

These normalizations were used to determine the images shown in Figure 4 in the main text. The extraction of the isolated tensor components in Figure 5 is then done using the relations shown in Table 1 in the main text. The corresponding pre-factors (Fresnel factors and angle-dependent terms are listed in Table S2.

| | 0-Fold | 1-Fold |
|---|---|---|
| $\chi^{(2),S}_{PPP,eff}(f_n)$ | $-0.203\chi^{(2)}_{xxz} + 0.107\chi^{(2)}_{zzz}$ | $-0.457\chi^{(2)}_{xxx} + 0.241\chi^{(2)}_{zzx}$ |
| $\chi^{(2),S}_{SSP,eff}(f_n)$ | $0.296\chi^{(2)}_{yyz}$ | $0.667\chi^{(2)}_{yyx}$ |
| $\chi^{(2),S}_{SSS,eff}(f_n)$ | - | $0.714\chi^{(2)}_{yyy}$ |

*Table S2. Pre-factors for the isolation of the different tensor components.*

### *SVD Analysis of Hyperspectral Images*

The SVD analysis of the hyperspectral images is performed to reduce the dimensions of the measured data for better visualization of the results. Additionally, this process also improves the data quality by reducing noise. In the first step of the analysis the normalized complex hyperspectral images are truncated along the frequency axis to isolate the relevant frequency range, followed by transforming the complex frequency data into the time domain using an inverse Fourier transformation. This yields a real valued 3D dataset (x, y, t) which is then analyzed using SVD. This results in a pair of data matrices, one 2D matrix containing the vibrational spectra (after Fourier transformation of the time domain data back into the frequency domain) of the different principal components, $\xi_n$, and a 3D matrix (x, y, n) that represents the amplitude of the of the different principal component spectra (n) as function of pixel position (x, y).

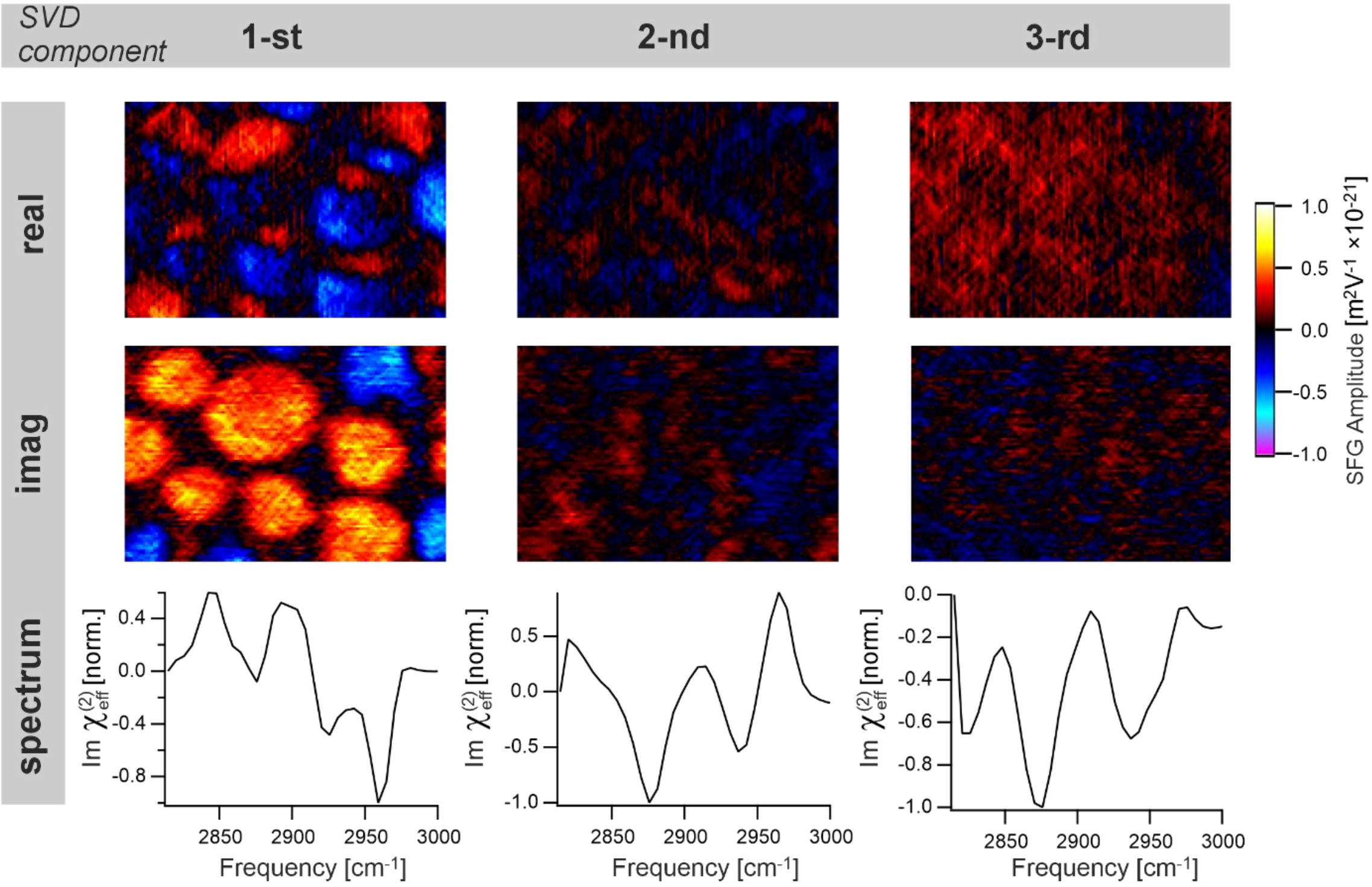

*Figure S1. SVD analysis of the 1-fold responses from the rotational measurements under the PPP polarization combination. Top and middle rows: amplitudes (absolute units) of the different SVD components (1st to 3rd SVD component) for real (top) and imaginary parts (middle) of the 1-fold rotational Fourier transformation. Bottom row: normalized spectra of the corresponding SVD components.*

For the 1-fold rotational data there are two separate 3D datasets of complex hyperspectral images, one representing the real and the other the imaginary part of the rotational Fourier transformation. For those cases a combined SVD analysis of both datasets is performed which consequently yields one set of vibrational spectra of the different principal components $\xi_n$, along with two 3D matrices (x, y, n) showing the contribution of the spectra for each pixel in a projection along the x (real part) and y axis (imaginary parts). An example of such analysis is presented in Figure S1, which shows the real and imaginary parts of the SVD amplitude images for the first three SVD components along with the corresponding spectra, all calculated for the 1-fold rotational frequency of the measurement under the PPP polarization combination. The data clearly shows that a contrast between the LC and LE phases can only be seen in the 1st SVD component. Although all three obtained spectra seem somewhat reasonable in terms of resonance peaks, the corresponding amplitude images for the 2nd and 3rd components show considerably smaller values and are dominated by noise. This demonstrates that the overall 1-fold data can be, to a very high accuracy, described by taking only the 1st SVD component. The same was found for all other rotational frequencies and polarization combinations, which is why only the 1st SVD component is shown in Figures 4, 5, and 7 in the main text.

In a second step, the real and imaginary amplitude images are transformed into magnitude and phase images as shown in Figure S2. This representation is very convenient as the magnitude image now represents the overall amount of 1-fold signal for each pixel while the corresponding phase image illustrates for which in-plane direction the corresponding SVD spectrum reaches its maximum, which is directly correlated with the molecular in-plane orientation. This representation in polar coordinates is used for all 1-fold responses shown in the main text.

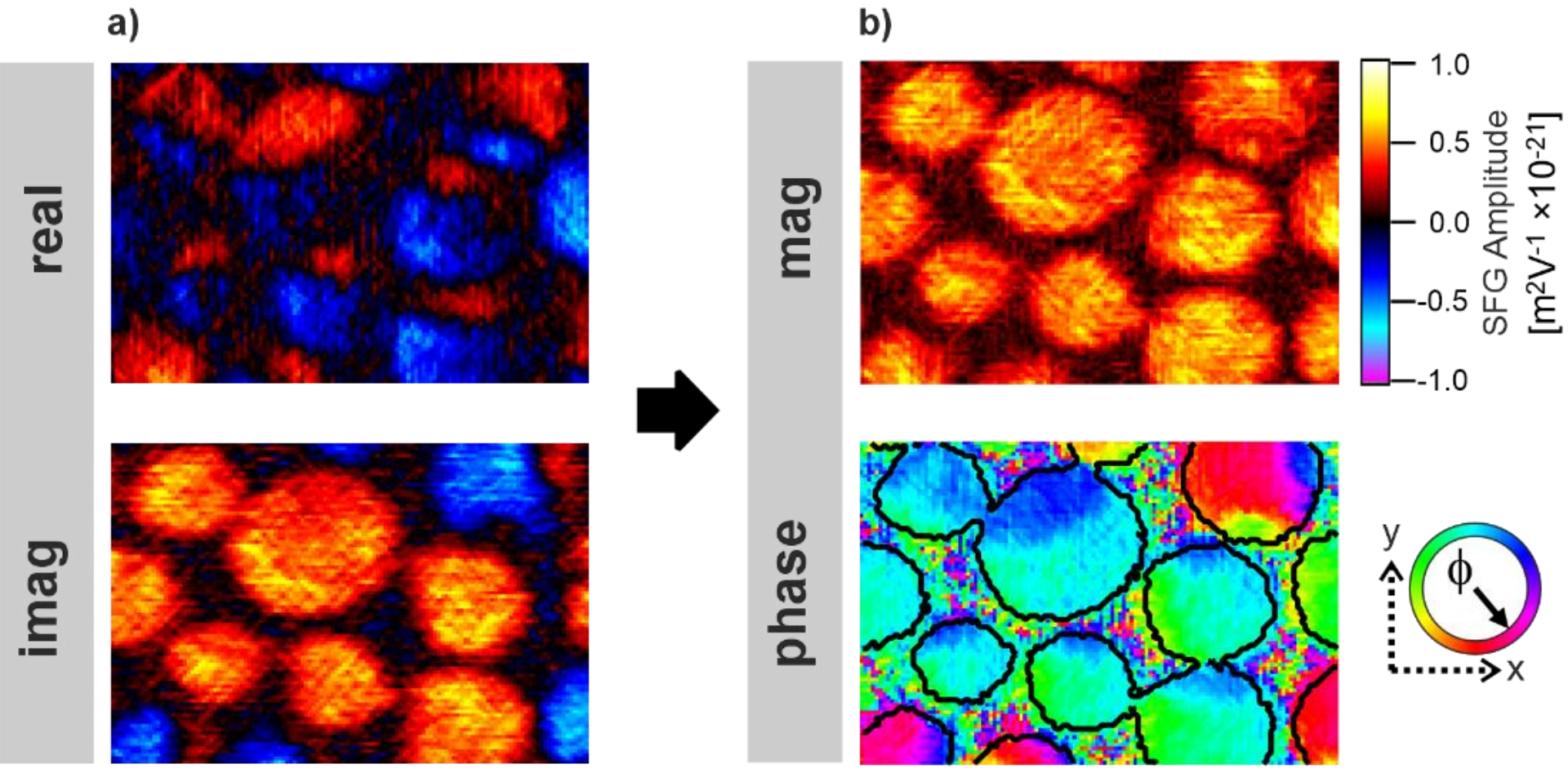

*Figure S2. Transformation of real and imaginary SVD images into polar coordinates.*

### *Comparison of In-Plane Phase Images from Different Tensor Components*

The 1-fold phase image shown in Figure S2 is obtained from the rotational measurements under PPP polarization combination. Similar phase images can obviously be extracted from the rotational measurements under SSS and SSP polarization presented in the main text. Furthermore, based on the following expressions in Eqs. S14-S16 for the 1-fold contributions of the methyl symmetric stretch to each polarization combination, the three phase maps should be closely related.

$$\chi^{(2)}_{PPP}(r^+,f_1) = \cos\theta'\, L_x(\omega_1)\rho\beta_{ccc}\sin\theta\left[-\frac{1}{4}\cos^2\theta'\, L_x(\omega_3)L_x(\omega_2)[R(1+3\cos^2\theta) +3\sin^2\theta] + \sin^2\theta'\, L_z(\omega_3)L_z(\omega_2)[R\sin^2\theta+\cos^2\theta]\right]e^{-i\phi} \quad (S14)$$

$$\chi^{(2)}_{SSP}(r^+,f_1) = \frac{1}{4}\cos\theta'\, L_y(\omega_3)L_y(\omega_2)L_x(\omega_1)\rho\beta_{ccc}\sin\theta\,[R(3+\cos^2\theta)+\sin^2\theta]e^{-i\phi} \quad (S15)$$

$$\chi^{(2)}_{SSS}(r^+,f_1) = \frac{i}{4}L_y(\omega_3)L_y(\omega_2)L_y(\omega_1)\rho\beta_{ccc}\sin\theta\,[R(1+3\cos^2\theta)+3\sin^2\theta]e^{-i\phi} \quad (S16)$$

Specifically, it is trivial to show that the rotational phases are related by the expressions in Eq. S17, where $n \in \{0,1\}$ is dependent on whether the PPP signal is dominated by the XXX ($n = 1$) or the ZZX ($n = 0$) susceptibility elements.

$$\arg\left(\chi^{(2)}_{PPP}(r^+,f_1)\right) + n\pi = \arg\left(\chi^{(2)}_{SSP}(r^+,f_1)\right) = \arg\left(\chi^{(2)}_{SSS}(r^+,f_1)\right) - \frac{\pi}{2} \quad (S17)$$

Figure S3 shows the phase images of the 1-fold contributions (calculated from SVD components, see above) for each of the PPP, SSP, and SSS polarization combinations, having rotated them according to Eq. S17 for better comparability. Specifically, we found that the XXX component dominates the 1-fold signal (which might be somewhat expected due to the higher experimental pre-factor compared to ZZX, see Table S2 above) and so rotated the phases of SSP and SSS based on $n = 1$.

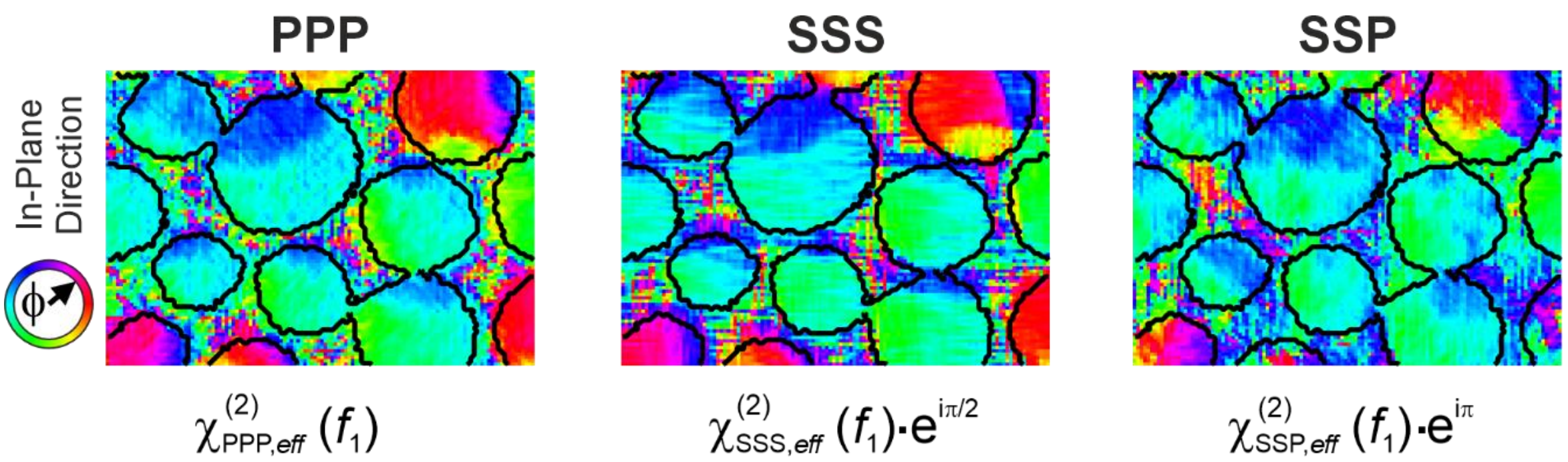

*Figure S3. Phase images showing the molecular in-plane direction of DPPC. Phase images are obtained from the rotational measurements under PPP, SSS, and SSP polarization combinations, respectively. Phase images for SSS and SSP polarizations are phase shifted by the theoretical values for better comparison.*

Overall, the phase images in Figure S3 shows exceptional comparability between the different polarization combinations. While this is to be expected as they are essentially showing the same maps i.e. maps of the in-plane azimuth angle, $\phi$, it also highlights the accuracy in determining this orientational parameter and shows that the sample seems stable over the entire duration of the experiments.

### *Fit Parameters from the Global Fit*

For the analysis presented in the main text, the different polarization combinations (PPP, SSP, and SSS) were used to deconvolute their individual contributing tensor elements, namely YYX, XXZ, ZZZ, ZZX, and XXX. These components are presented in Figure 5 in the main text in the form of a 2D image in absolute units and a normalized spectrum (calculated using SVD, see above). In order to extract structural information from these data, the spectral line-shapes were compared and the absolute amplitudes of the methyl symmetric stretch were determined. For this, we utilized a Lorentzian global fitting procedure that fitted both the real and imaginary parts of all five spectra using linked resonant parameters. Specifically, the center frequencies and damping coefficients were shared across all spectra and the resonant amplitudes linked between the real and imaginary parts of each spectrum. A summary of the fitting parameters that correspond to the fitted spectra shown in Figure 5b in the main text is given below in Table S3. It is important to note that, as the spectra are normalized, the fitted Lorentzian amplitudes, $A$, have the same units as the damping constants, $\Gamma$, i.e. cm$^{-1}$. For better clarity, we have instead presented the spectral amplitudes of the fitted resonances, i.e. the dimensionless quantity $\frac{A}{\Gamma}$ that reflects the amplitude reached by the resonance at its center frequency.

| Frequency, $\omega$ / cm$^{-1}$ | Damping Constant, $\Gamma$ / cm$^{-1}$ | Spectral Amplitude, $\frac{A}{\Gamma}$ | | | | |
|---|---|---|---|---|---|---|
| | | XXX | XXZ | YYX | ZZX | ZZZ |
| 2847.7 | 8.2 | 0.639 | -0.063 | 0.860 | 0.098 | -0.127 |
| 2879.3 | 10.3 | -0.782 | -1.129 | -0.557 | -1.242 | -0.928 |
| 2889.2 | 11.5 | 0.832 | 0.108 | -0.619 | 0.196 | 0.009 |
| 2911.9 | 9.9 | 0.549 | 0.084 | 0.671 | 0.322 | -0.157 |
| 2923.1 | 9.8 | -0.723 | -0.063 | 0.291 | -0.097 | -0.271 |
| 2939.8 | 9.6 | -0.214 | -0.750 | -0.604 | -0.547 | -0.684 |

| 2959.9 | 6.5 | -0.924 | 0.411 | -0.090 | 0.111 | -0.144 |
|---|---|---|---|---|---|---|
| 2966.6 | 5.5 | -0.184 | 0.306 | 0.386 | 0.207 | -0.898 |
| Real Offset | | 0.047 | 0.169 | 0.024 | 0.285 | 0.456 |

*Table S3: Fitting parameters that resulted from the global fitting procedure on the five extracted tensor elements.*

***Calculation of Molecular Densities***

In the main text, we showed that the different tensor elements are all equally scaled by the molecular surface density of the species being probed, $\rho$ (see for example Eqs. 15-19). Therefore, with the extracted tilt angle, one can remove the orientational dependency of the 0-fold contributions to get an image that reflects only the density changes. In principle this could be done with any individual tensor element (also considering the cancellation parameter for the 1-fold components) or the overall response in a given polarization combination. Here, we used the 0-fold PPP response.

While the resulting images from removing the orientational dependency reflect the local density fluctuations of the molecules being probed, they are not correctly scaled as absolute densities, despite the susceptibilities being in absolute units. This is due to each tensor element being scaled by $\rho\beta_{ccc}$ (i.e. the product of the density and a single element of the molecular hyperpolarizability), and thus the scaling by the hyperpolarizability would have to be removed. Although this parameter could be calculated using literature values and the bond additivity model(*97*, *135*), this is not necessary. Instead, by integrating the orientation-corrected images and dividing by the surface area, the result must yield the average density (assuming of course that the spatial averaging is over a sufficiently large enough area), $\langle\rho\rangle$, which is a parameter that is easily extracted from the surface pressure (Langmuir) isotherm. Therefore, the absolute densities can be extracted as in Eq. S18,

$$\rho(x,y) = \frac{\left.\dfrac{\chi^{(2)}_{PPP,eff}(f_0)}{\Omega(\theta)}\right|_{x,y}}{\left\langle\dfrac{\chi^{(2)}_{PPP,eff}(f_0)}{\Omega(\theta)}\right\rangle}\langle\rho\rangle \qquad (S18)$$

where $\Omega(\theta)$ is the orientational part of the total 0-fold PPP response (i.e. also including the Fresnel factors and $R$).

In this work, the films here were compressed to a surface pressure of 20 $mNm^{-1}$, which corresponds to an average density of 2.5 $nm^{-2}$. Therefore, with the composition ratio of 4:1 DPPC:dPOPC, the DPPC density can be calculated from the C-H PPP 0-fold response using $\langle\rho\rangle = 2.0\,nm^{-2}$, and the corresponding dPOPC density calculated from the C-D PPP 0-fold response using $\langle\rho\rangle = 0.5\,nm^{-2}$.